%% file: paper.tex
\documentclass{sig-alternate}

\usepackage{import}

\usepackage{caption}
\usepackage{subcaption}

\usepackage{algorithmic}
\usepackage{algorithm}

\usepackage{array}

\usepackage{graphicx}
\usepackage{epstopdf}

\usepackage{color}

\usepackage[english]{babel}
\usepackage{adjustbox}
\usepackage{color, colortbl}
\definecolor{high}{rgb}{0.7,.85,.85}
\usepackage{cite}
\usepackage{mathtools}

\usepackage{enumitem}

\DeclarePairedDelimiter{\ceil}{\lceil}{\rceil}

\graphicspath{{./figures/overview/}}

\newcommand{\blogcatalog}{\textsc{BlogCatalog}}
\newcommand{\flickr}{\textsc{Flickr}}

\newcommand{\youtube}{\textsc{YouTube}}

\newcommand{\todo}[1]{{\small\color{red}{\bf #1}}}

\newcommand{\ouralgorithm}{\textsc{DeepWalk}}

\newcommand{\socdimL}{\text{SpectralClustering}}
\newcommand{\socdimB}{\text{Modularity}}
\newcommand{\socdimA}{\text{EdgeCluster}}
\newcommand{\wvrn}{\text{wvRN}}

\newcommand{\Laplacian}{\widetilde{\mathcal{L}}}
\newcommand{\comment}[1]{}

\begin{document}
%

\title{DeepWalk: 
Online Learning of Social Representations
}

%
%
%
%
%

\numberofauthors{3} 
%
\author{
%
%
\alignauthor
       Bryan Perozzi\\
\affaddr{Stony Brook University}\\
\affaddr{Department of Computer Science}\\
\alignauthor
	   Rami Al-Rfou\\
\affaddr{Stony Brook University}\\
\affaddr{Department of Computer Science}\\
\alignauthor 
	   Steven Skiena\\
\affaddr{Stony Brook University}\\
\affaddr{Department of Computer Science}\\
\end{tabular}\newline\begin{tabular}{c}
\{bperozzi, ralrfou, skiena\}@cs.stonybrook.edu\\
}
\date{23 August 2014}

\maketitle
\begin{abstract}
We present \ouralgorithm, a novel approach for learning latent representations of vertices in a network. 
These latent representations encode social relations in a continuous vector space, which is easily exploited by statistical models.
\ouralgorithm\ generalizes recent advancements in language modeling and unsupervised feature learning  (or \emph{deep learning}) from sequences of words to graphs.

\ouralgorithm\ uses local information obtained from truncated random walks to \emph{learn} latent representations by treating walks as the equivalent of sentences.
We demonstrate  \ouralgorithm's latent representations on several multi-label network classification tasks for social networks such as BlogCatalog, Flickr, and YouTube.  
Our results show that \ouralgorithm\ outperforms challenging baselines which are allowed a global view of the network, especially in the presence of missing information. 
\ouralgorithm's representations can provide $F_1$ scores up to 10\% higher than competing methods when labeled data is sparse.
In some experiments, \ouralgorithm's representations are able to outperform all baseline methods while using 60\% less training data.


\ouralgorithm\ is also scalable.  
It is an online learning algorithm which builds useful incremental results, and is trivially parallelizable.
These qualities make it suitable for a broad class of real world applications such as network classification, and anomaly detection.
\end{abstract}





\section{Introduction}

The sparsity of a network representation is both a strength and a weakness.
Sparsity enables the design of efficient discrete algorithms, but can make it harder to generalize in statistical learning.
Machine learning applications in networks (such as network classification \cite{getoor2007introduction,sen2008collective}, content recommendation \cite{fouss2007random}, anomaly detection \cite{chandola2009anomaly}, and missing link prediction \cite{liben2007link}) must be able to deal with this sparsity in order to survive.

In this paper we introduce \emph{deep learning} (unsupervised feature learning) \cite{deepfuture} techniques, which have proven successful in natural language processing, into network analysis for the first time.
We develop an algorithm (\ouralgorithm) that learns \emph{social representations} of a graph's vertices, by modeling a stream of short random walks.
Social representations are latent features of the vertices that capture neighborhood similarity and community membership.
These latent representations encode social relations in a continuous vector space with a relatively small number of dimensions.
\ouralgorithm\ generalizes neural language models to process a special language composed of a set of randomly-generated walks.
These neural language models have been used to capture the semantic and syntactic structure of human language\cite{senna1}, and even logical analogies \cite{regularities}.

\begin{figure}[t!]
	\centering
        \begin{subfigure}[b]{0.48\columnwidth}
                \includegraphics[width=\columnwidth]{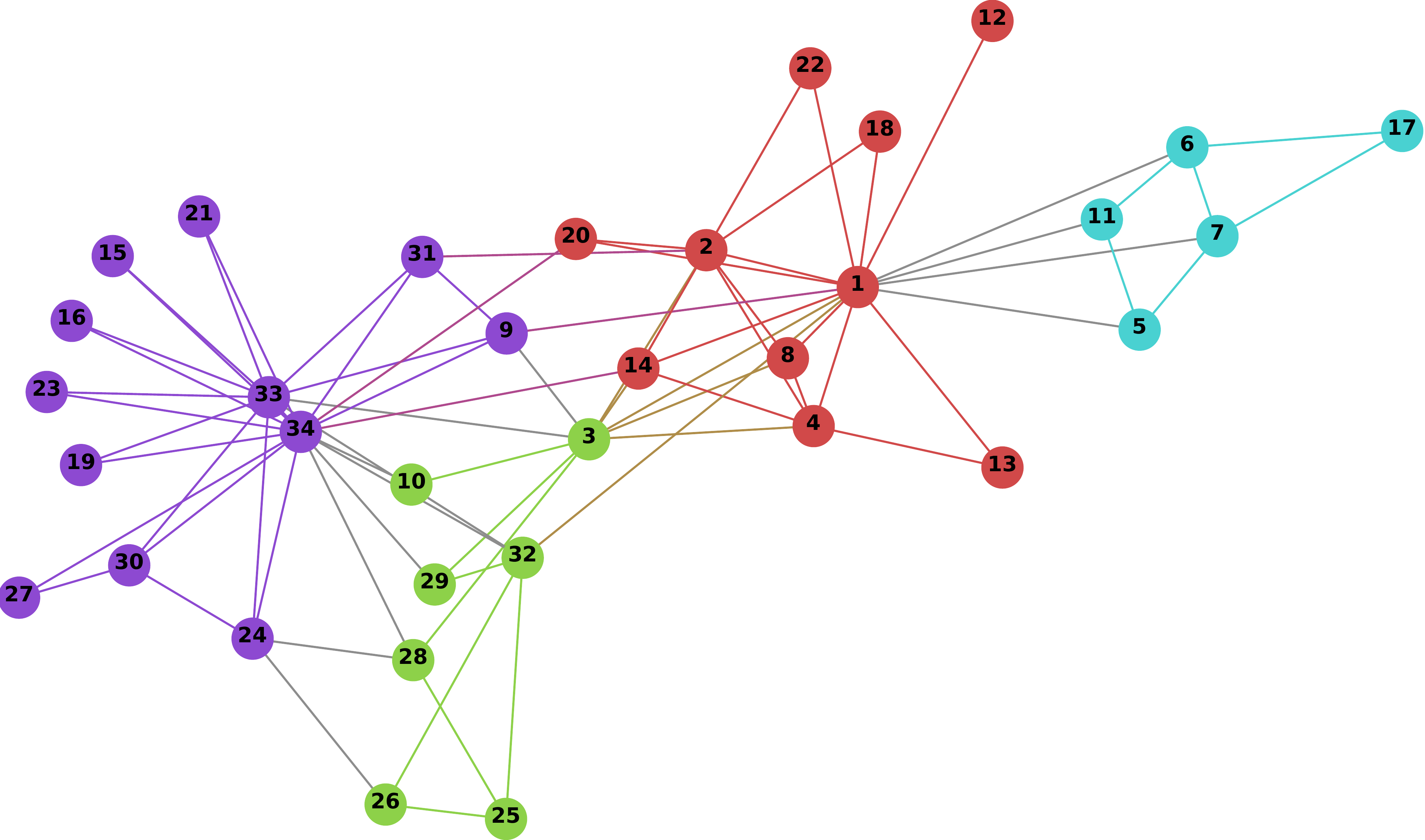}
                \caption{Input: Karate Graph}
                \label{fig:toy_example_graph}
        \end{subfigure}
        \begin{subfigure}[b]{0.48\columnwidth}
                \includegraphics[width=\columnwidth]{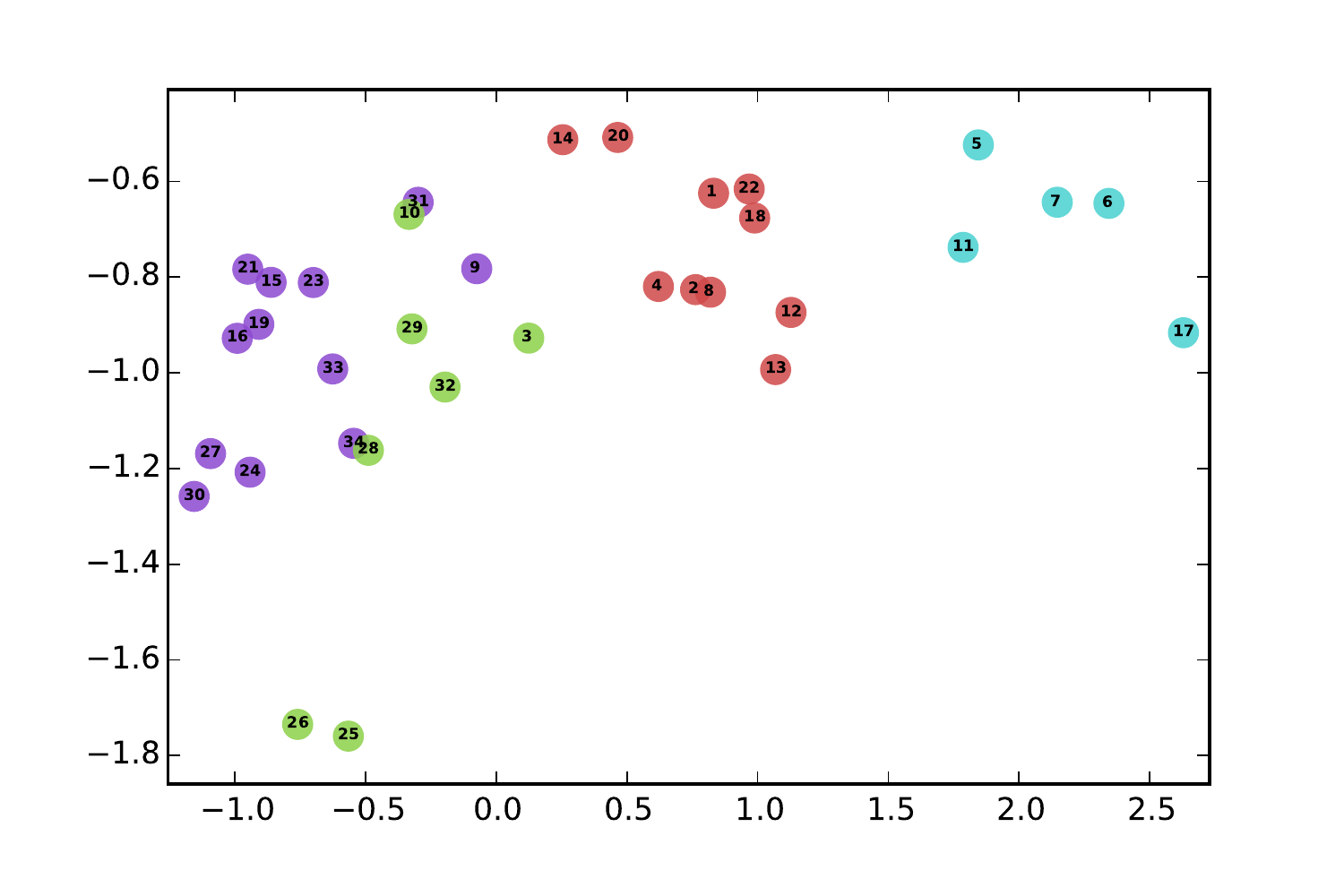}
                \caption{Output: Representation}
                \label{fig:toy_example_embedding}
        \end{subfigure}
        \caption{Our proposed method \emph{learns} a latent space representation of social interactions in $\mathbb{R}^d$.  The learned representation encodes community structure so it can be easily exploited by standard classification methods. Here, our method is used on Zachary's Karate network \cite{zachary1977information} to generate a latent representation in $\mathbb{R}^2$.  
		Note the correspondence between community structure in the input graph and the embedding. Vertex colors represent a modularity-based clustering of the input graph.
        }
        \label{fig:toy_example}
\end{figure}

\ouralgorithm\ takes a graph as input and produces a latent representation as an output.
The result of applying our method to the well-studied Karate network is shown in Figure \ref{fig:toy_example}.  
The graph, as typically presented by force-directed layouts, is shown in Figure \ref{fig:toy_example_graph}.
Figure \ref{fig:toy_example_embedding} shows the output of our method with 2 latent dimensions.
Beyond the striking similarity, we note that linearly separable portions of (\ref{fig:toy_example_embedding}) correspond to clusters found through modularity maximization in the input graph (\ref{fig:toy_example_graph}) (shown as vertex colors).

To demonstrate \ouralgorithm's potential in real world scenarios, we evaluate its performance on challenging multi-label network classification problems in large heterogeneous graphs.
In the relational classification problem, the links between feature vectors violate the traditional \emph{i.i.d.} assumption.  
Techniques to address this problem typically use approximate inference techniques \cite{neville2000iterative,Pearl:1988:PRI:534975} to leverage the dependency information to improve classification results.
We distance ourselves from these approaches by learning label-independent representations of the graph.
Our representation quality is not influenced by the choice of labeled vertices, so they can be shared among tasks.

\ouralgorithm\ outperforms other latent representation methods for creating \emph{social dimensions} \cite{Tang:2009:RLV:1557019.1557109,Tang:2011:Leveraging}, especially when labeled nodes are scarce.
Strong performance with our representations is possible with very simple linear classifiers (e.g. logistic regression).
Our representations are general, and can be combined with any classification method (including iterative inference methods).
\ouralgorithm\ achieves all of that while being an online algorithm that is trivially parallelizable.

Our contributions are as follows: 

\begin{itemize}

\item We introduce deep learning as a tool to analyze graphs, to build robust representations that are suitable for statistical modeling. \ouralgorithm\ learns structural regularities present within short random walks.

\item We extensively evaluate our representations on multi-label classification tasks on several social networks.
We show significantly increased classification performance in the presence of label sparsity, getting improvements 5\%-10\% of Micro $F_1$, on the sparsest problems we consider.
In some cases, \ouralgorithm's representations can outperform its competitors even when given 60\% less training data.

\item We demonstrate the scalability of our algorithm by building representations of web-scale graphs, (such as YouTube) using a parallel implementation.
Moreover, we describe the minimal changes necessary to build a streaming version of our approach.

\end{itemize}

The rest of the paper is arranged as follows.  In Sections \ref{sec:problem} and \ref{sec:SRL}, we discuss the problem formulation of classification in data networks, and how it relates to our work.  In Section \ref{sec:method} we present \ouralgorithm, our approach for Social Representation Learning.  We outline ours experiments in Section \ref{sec:experimental_design}, and present their results in Section \ref{sec:experiments}.  We close with a discussion of related work in Section \ref{sec:related}, and our conclusions.

\section{Problem Definition}
\label{sec:problem}

We consider the problem of classifying members of a social network into one or more categories.  
More formally, let $G=(V, E)$, where $V$ are the members of the network, and $E$ be its edges, $E \subseteq (V \times V)$.
Given a partially labeled social network $G_L = (V,E,X,Y)$, with attributes $X \in \mathbb{R}^{|V|\times S}$ where $S$ is the size of the feature space for each attribute vector, and $Y \in \mathbb{R}^{|V|\times |\mathcal{Y}|}$, $\mathcal{Y}$ is the set of labels.

In a traditional machine learning classification setting, we aim to learn a hypothesis $H$ that maps elements of $X$ to the labels set $\mathcal{Y}$.
In our case, we can utilize the significant information about the dependence of the examples embedded in the structure of $G$ to achieve superior performance.

In the literature, this is known as the relational classification (or the \emph{collective classification} problem \cite{sen2008collective}).
Traditional approaches to relational classification pose the problem as an inference in an undirected Markov network, and then use iterative approximate inference algorithms (such as the iterative classification algorithm \cite{neville2000iterative}, Gibbs Sampling \cite{geman1984stochastic}, or label relaxation \cite{hummel1983foundations}) to compute the posterior distribution of labels given the network structure.

We propose a different approach to capture the network topology information.
Instead of mixing the label space as part of the feature space, we propose an unsupervised method which learns features that capture the graph structure \emph{independent} of the labels' distribution.

This separation between the structural representation and the labeling task avoids cascading errors, which can occur in iterative methods \cite{neville2008bias}.
Moreover, the same representation can be used for multiple classification problems concerning that network.

Our goal is to learn $X_E \in \mathbb{R}^{|V|\times d}$, where $d$ is small number of latent dimensions.
These low-dimensional representations are distributed; meaning each social phenomena is expressed by a subset of the dimensions and each dimension contributes to a subset of the social concepts expressed by the space.

Using these structural features, we will augment the attributes space to help the classification decision.  
These features are general, and can be used with any classification algorithm (including iterative methods).  
However, we believe that the greatest utility of these features is their easy integration with simple machine learning algorithms. They scale appropriately in real-world networks, as we will show in Section \ref{sec:experiments}.








\section{Learning Social Representations}
\label{sec:SRL}


We seek learning social representations with the following characteristics:
\begin{itemize}
\item \textbf{Adaptability} - Real social networks are constantly evolving;  new social relations should not require repeating the learning process all over again.

\item \textbf{Community aware} - The distance between latent dimensions should represent a metric for evaluating social similarity between the corresponding members of the network.
This allows generalization in networks with homophily.



\item \textbf{Low dimensional} - When labeled data is scarce, low-dimensional models generalize better, and speed up convergence and inference.

\item \textbf{Continuous} - 
We require latent representations to model partial community membership in continuous space.
In addition to providing a nuanced view of community membership, a continuous representation has smooth decision boundaries between communities which allows more robust classification.



\end{itemize}

Our method for satisfying these requirements learns representation for vertices from a stream of short random walks, using optimization techniques originally designed for language modeling.
Here, we review the basics of both random walks and language modeling, and describe how their combination satisfies our requirements.

\subsection{Random Walks}
We denote a random walk rooted at vertex $v_i$ as $\mathcal{W}_{v_i}$.  
It is a stochastic process with random variables $\mathcal{W}^1_{v_i},\mathcal{W}^2_{v_i},\dots{},\mathcal{W}^k_{v_i}$ such that $\mathcal{W}^{k+1}_{v_i}$ is a vertex chosen at random from the neighbors of vertex $v_k$.
Random walks have been used as a similarity measure for a variety of problems in content recommendation \cite{fouss2007random} and community detection \cite{andersen2006local}.
They are also the foundation of a class of \emph{output sensitive} algorithms which use them to compute local community structure information in time sublinear to the size of the input graph \cite{spielman2004nearly}.

It is this connection to local structure that motivates us to use a \emph{stream} of short random walks as our basic tool for extracting information from a network.
In addition to capturing community information, using random walks as the basis for our algorithm gives us two other desirable properties.  First, 
local exploration is easy to parallelize.  Several random walkers (in different threads, processes, or machines) can simultaneously explore different parts of the same graph.
Secondly, relying on information obtained from short random walks make it possible to accommodate small changes in the graph structure without the need for global recomputation.
We can iteratively update the learned model with new random walks from the changed region in time sub-linear to the entire graph.


\subsection{Connection: Power laws}
Having chosen online random walks as our primitive for capturing graph structure, we now need a suitable method to capture this information.  
If the degree distribution of a connected graph follows a power law (is \emph{scale-free}), we observe that the frequency which vertices appear in the short random walks will also follow a power-law distribution.  

Word frequency in natural language follows a similar distribution, and techniques from language modeling account for this distributional behavior.
To emphasize this similarity we show two different power-law distributions in Figure \ref{fig:power_law}. 
The first comes from a series of short random walks on a scale-free graph, and the second comes from the text of 100,000 articles from the English Wikipedia.

A core contribution of our work is the idea that techniques which have been used to model natural language (where the symbol frequency follows a power law distribution (or \emph{Zipf's law})) can be re-purposed to model community structure in networks.
\begin{figure}
	\centering
      \begin{subfigure}[b]{0.48\columnwidth}
                \includegraphics[width=\columnwidth]{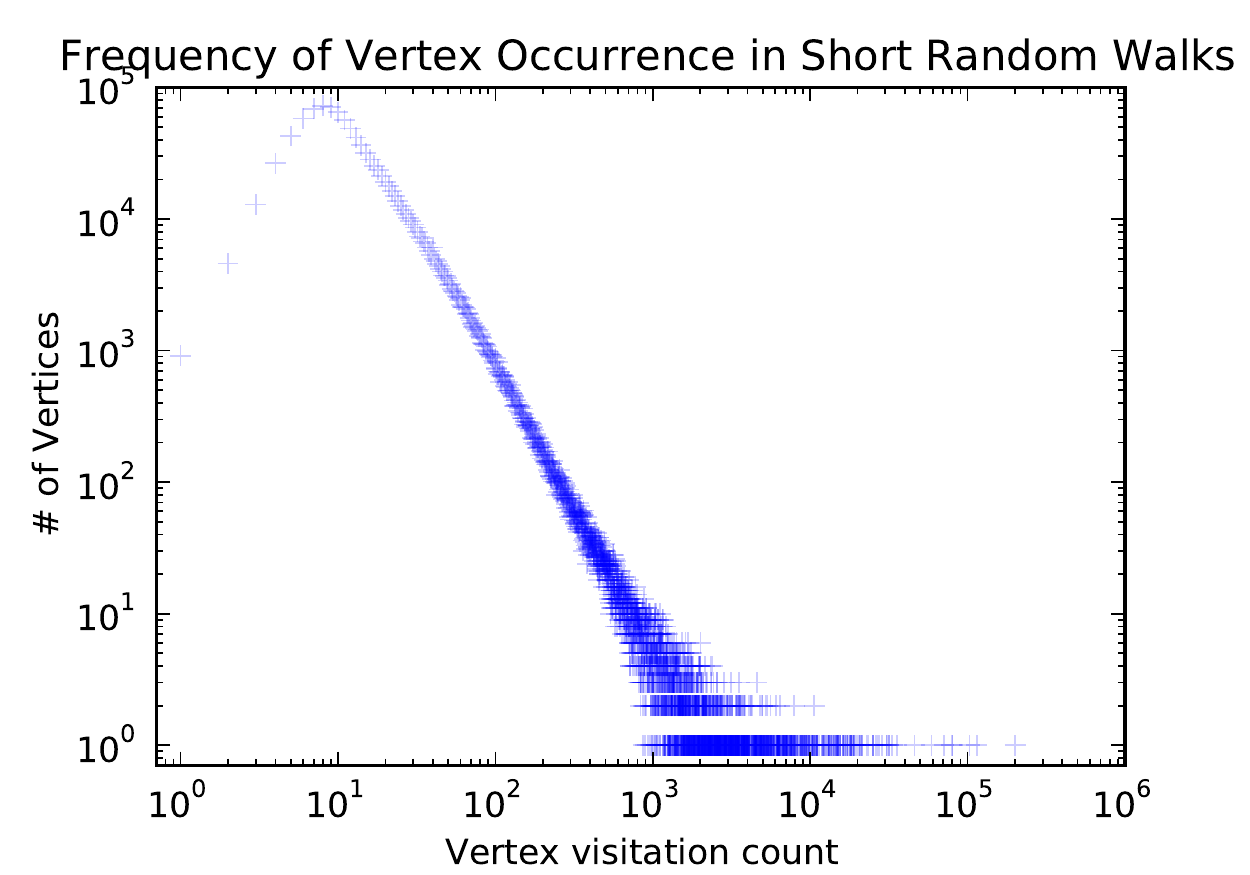}
		\caption{YouTube Social Graph}
		\label{fig:powerlaw-youtube}
	\end{subfigure}
      \begin{subfigure}[b]{0.48\columnwidth}
                \includegraphics[width=\columnwidth]{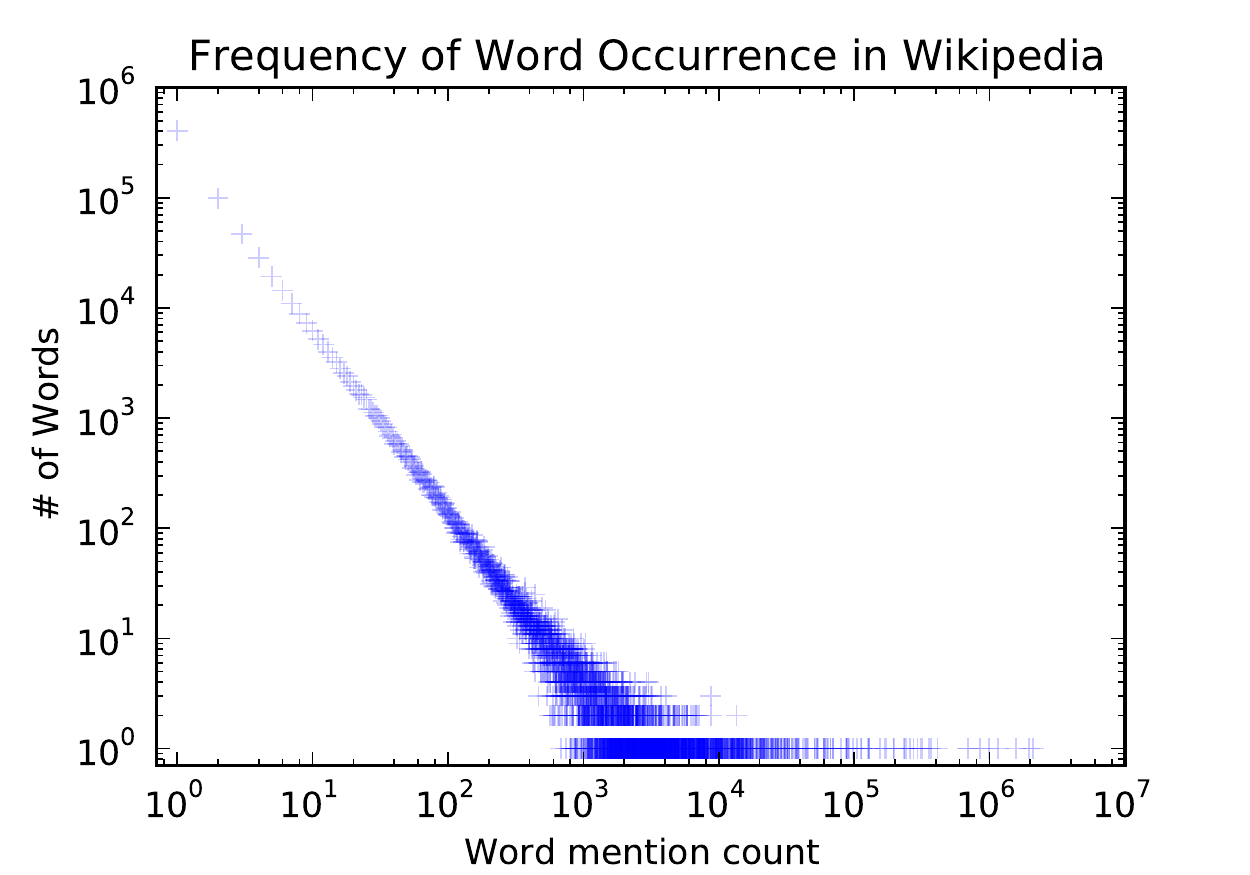}	
		\caption{Wikipedia Article Text}
		\label{fig:powerlaw-wiki}
	\end{subfigure}
\caption{The power-law distribution of vertices appearing in short random walks (\ref{fig:powerlaw-youtube}) follows a power-law, much like the distribution of words in natural language (\ref{fig:powerlaw-wiki}).  
}
\label{fig:power_law}
\end{figure}
We spend the rest of this section reviewing the growing work in language modeling, and transforming it to learn representations of vertices which satisfy our criteria.


\subsection{Language Modeling}
The goal of language modeling is estimate the likelihood of a specific sequence of words appearing in a corpus.
More formally, given a sequence of words $$W_{1}^{n} = (w_0, w_1, \cdots, w_n)$$ where $w_i \in \mathcal{V}$ ($\mathcal{V}$ is the vocabulary), we would like to maximize the $\Pr(w_n| w_0, w_1, \cdots, w_{n-1})$ over all the training corpus.
\comment{
The performance of the model will be evaluated on held out testing corpus using a perplexity score $PP$ as defined below
\begin{equation}
PP = \frac{1}{\sqrt[n]{P(W_1^n)}}
\end{equation}}

Recent work in representation learning has focused on using probabilistic neural networks to build general representations of words which extend the scope of language modeling beyond its original goals.

In this work, we present a generalization of language modeling to explore the graph through a stream of short random walks.
These walks can be thought of short sentences and phrases in a special language.
The direct analog is to estimate the likelihood of observing vertex $v_i$ given all the previous vertices visited so far in the random walk.

$$\Pr\big(v_{i}\mid( v_1, v_2, \cdots, v_{i-1})\big)$$

Our goal is to learn a latent representation, not only a probability distribution of node co-occurrences, and so we introduce a mapping function $\Phi \colon v \in V \mapsto \mathbb{R}^{|V|\times d}$.
This mapping $\Phi$ represents the latent social representation associated with each vertex $v$ in the graph.
(In practice, we represent $\Phi$ by a $|V| \times d$ matrix of free parameters, which will serve later on as our $X_E$.)
The problem then, is to estimate the likelihood:

\begin{equation}
\Pr\Big(v_i \mid \big(\Phi(v_1), \Phi(v_2), \cdots, \Phi(v_{i-1})\big)\Big)
\end{equation}

However as the walk length grows, computing this objective function becomes unfeasible.

A recent relaxation in language modeling \cite{word2vec1,word2vec2} turns the prediction problem on its head.
First, instead of using the context to predict a missing word, it uses one word to predict the context.
Secondly, the context is composed of the words appearing to right side of the given word as well as the left side.
Finally, it removes the ordering constraint on the problem.
Instead, the model is required to maximize the probability of any word appearing in the context without the knowledge of its offset from the given word.

In terms of vertex representation modeling, this yields the optimization problem:

\begin{equation}
\begin{aligned}
& \underset{\Phi}{\text{minimize}}
& & -\log \Pr\big(\{v_{i-w}, \cdots, v_{i-1}, v_{i+1}, \cdots, v_{i+w}\}\mid \Phi(v_i) \big) \\
\end{aligned}
\label{eq:objective}
\end{equation}

We find these relaxations are particularly desirable for social representation learning.
First, the order independence assumption better captures a sense of `nearness' that is provided by random walks.
Moreover, this relaxation is quite useful for speeding up the training time by building small models as one vertex is given at a time.

Solving the optimization problem from Eq. \ref{eq:objective} builds representations that capture the shared similarities in local graph structure between vertices.
Vertices which have similar neighborhoods will acquire similar representations (encoding co-citation similarity), and allowing generalization on machine learning tasks.

By combining both truncated random walks and neural language models we formulate a method which satisfies all of our desired properties.
This method generates representations of social networks that are low-dimensional, and exist in a continuous vector space.
Its representations encode latent forms of community membership, and because the method outputs useful intermediate representations, it can adapt to changing network topology.

\section{Method}
\label{sec:method}

\begin{figure*}

\begin{subfigure}[b]{0.30\textwidth}
\adjustbox{trim={0.0\width} {.0\height} {0.7\width} {.05\height},clip}{\includegraphics{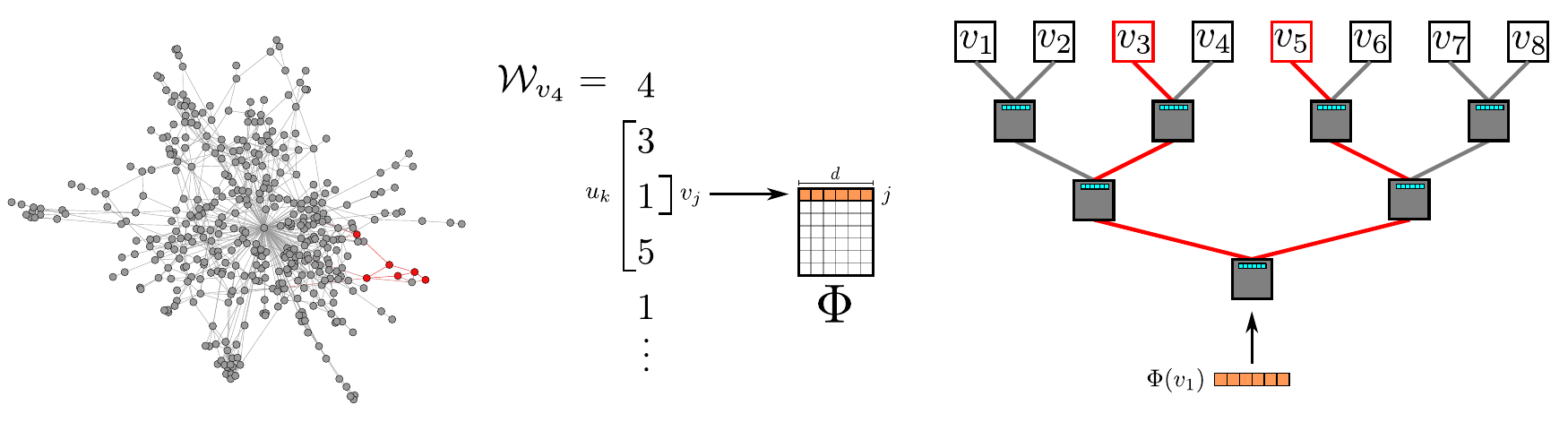}}
\caption{Random walk generation.}
\label{fig:graph}
\end{subfigure}
\begin{subfigure}[b]{0.30\textwidth}
\adjustbox{trim={.3\width} {.15\height} {0.4\width} {.05\height},clip}{\includegraphics{figures/deepwalk_overview}}
\caption{Representation mapping.}
\label{fig:phi}
\end{subfigure}
\begin{subfigure}[b]{0.4\textwidth}
\adjustbox{trim={.605\width} {.1\height} {0.0\width} {.04\height},clip}{\includegraphics{figures/deepwalk_overview}}
\caption{Hierarchical Softmax.}
\label{fig:hsm}
\end{subfigure}

\caption{Overview of \ouralgorithm.
We slide a window of length $2w+1$ over the random walk $\mathcal{W}_{v_4}$, mapping the central vertex $v_1$ to its representation $\Phi(v_1)$.
Hierarchical Softmax factors  out $\Pr(v_3 \mid \Phi(v_1))$ and $\Pr(v_5 \mid \Phi(v_1))$ over sequences of probability distributions corresponding to the paths starting at the root and ending at $v_3$ and $v_5$.
The representation $\Phi$ is updated to maximize the probability of $v_1$ co-occurring with its context $\{v_3, v_5\}$.
}
\end{figure*}

In this section we discuss the main components of our algorithm. 
We also present several variants of our approach and discuss their merits.

\subsection{Overview}
As in any language modeling algorithm, the only required input is a corpus and a vocabulary $\mathcal{V}$.
\ouralgorithm\  considers a set of short truncated random walks its own corpus, and the graph vertices as its own vocabulary ($\mathcal{V} = V$).
While it is beneficial to know the $V$ and the frequency distribution of vertices in the random walks ahead of the training, it is not necessary for the algorithm to work as we will show in \ref{sec:hsm}.

\subsection{Algorithm: {\large \ouralgorithm}}
The algorithm consists of two main components; first a random walk generator and second an update procedure.

\begin{algorithm}[t]
\begin{algorithmic}[1]
\REQUIRE graph $G(V,E)$\\ window size $w$\\ embedding size $d$\\ walks per vertex $\mathcal{\gamma}$ \\ walk length $t$
\ENSURE matrix of vertex representations $\Phi \in \mathbb{R}^{|V| \times d}$
	\STATE Initialization: Sample $\Phi$ from $\mathcal{U}^{|V| \times d}$
	\STATE Build a binary Tree $T$ from $V$
	\FOR{$i=0$ to $\mathcal{\gamma}$}
		\STATE	$\mathcal{O} = \text{Shuffle}(V)$
        \FOR{\textbf{each} $v_i \in \mathcal{O}$}
		\STATE $\mathcal{W}_{v_i} = RandomWalk(G, v_i, $t$) $
		\STATE SkipGram($\Phi$, $\mathcal{W}_{v_i}$, $w$)
		 \ENDFOR
	\ENDFOR
\end{algorithmic}
\caption{\ouralgorithm($G$, $w$, $d$, $\gamma$, $t$)}
\label{alg:deepwalk}
\end{algorithm}

The random walk generator takes a graph $G$ and samples uniformly a random vertex $v_i$ as the root of the random walk $\mathcal{W}_{v_i}$.
A walk samples uniformly from the neighbors of the last vertex visited until the maximum length ($t$) is reached.
While we set the length of our random walks in the experiments to be fixed, there is no restriction for the random walks to be of the same length.
These walks could have restarts (i.e. a teleport probability of returning back to their root), but
our preliminary results did not show any advantage of using restarts.
In practice, our implementation specifies a number of random walks $\mathcal{\gamma}$ of length $t$ to start at each vertex.

Lines 3-9 in Algorithm \ref{alg:deepwalk} shows the core of our approach.
The outer loop specifies the number of times, $\mathcal{\gamma}$, which we should start random walks at each vertex.
We think of each iteration as making a `pass' over the data and sample one walk per node during this pass.
At the start of each pass we generate a random ordering to traverse the vertices.  
This is not strictly required, but is well-known to speed up the convergence of stochastic gradient descent.

In the inner loop, we iterate over all the vertices of the graph.  
For each vertex $v_i$ we generate a random walk $|\mathcal{W}_{v_i}| = t$, and then use it to update our representations (Line 7).
We use the SkipGram algorithm \cite{word2vec1} to update these representations in accordance with our objective function in Eq. \ref{eq:objective}.

\subsubsection{SkipGram}
SkipGram is a language model that maximizes the co-occurrence probability among the words that appear within a window, $w$, in a sentence \cite{word2vec1}.

Algorithm \ref{alg:skipgram} iterates over all possible collocations in  random walk that appear within the window $w$ (lines 1-2).
For each, we map each vertex $v_j$ to its current representation vector $\Phi(v_j) \in \mathbb{R}^d$ (See Figure \ref{fig:phi}).
Given the representation of $v_j$, we would like to maximize the probability of its neighbors in the walk (line 3).
We can learn such posterior distribution using several choices of classifiers.
For example, modeling the previous problem using logistic regression would result in a huge number of labels  that is equal to $|V|$ which could be in millions or billions.
Such models require large amount of computational resources that could span a whole cluster of computers \cite{nnlm}.
To speed the training time, Hierarchical Softmax \cite{hsm1,hsm2} can be used to approximate the probability distribution.

\begin{algorithm}[t]
\begin{algorithmic}[1]
    \FOR{\textbf{each} $v_j \in \mathcal{W}_{v_i}$}
		\FOR{\textbf{each} $u_k \in \mathcal{W}_{v_i}[j-w: j+w]$}
				\STATE  $J(\Phi) = - \log{\Pr(u_k \mid \Phi(v_j))}$
	        	\STATE $\Phi = \Phi - \alpha * \frac{\partial J}{\partial \Phi}$
		\ENDFOR

	\ENDFOR

\end{algorithmic}
\caption{SkipGram($\Phi$, $\mathcal{W}_{v_i}$, $w$)}
\label{alg:skipgram}
\end{algorithm}

\subsubsection{Hierarchical Softmax}
\label{sec:hsm}
Given that $u_k \in V$, calculating $\Pr(u_k \mid \Phi(v_j))$ in line 3 is not feasible.
Computing the partition function (normalization factor) is expensive.
If we assign the vertices to the leaves of a binary tree, the prediction problem turns into maximizing the probability of a specific path in the tree (See Figure \ref{fig:hsm}).
If the path to vertex $u_k$ is identified by a sequence of tree nodes $(b_0, b_1, \dots, b_{\ceil{\log|V|}})$, ($b_0$ = root, $b_{\ceil{\log|V|}} = u_k$) then $$\Pr(u_k \mid \Phi(v_j)) = \prod_{l=1}^{\ceil{\log|V|}} \Pr(b_l \mid \Phi(v_j)) $$

Now, $\Pr(b_l \mid \Phi(v_j))$ could be modeled by a binary classifier that is assigned to the parent of the node $b_l$.
This reduces the computational complexity of calculating $\Pr(u_k \mid \Phi(v_j))$ from $O(|V|)$ to $O(\log |V|)$.

We can speed up the training process further, by assigning shorter paths to the frequent vertices in the random walks.
Huffman coding is used to reduce the access time of frequent elements in the tree.

\subsubsection{Optimization}
The model parameter set is $\{\Phi, T\}$ where the size of each is $O(d|V|)$.
Stochastic gradient descent (SGD) \cite{sgd} is used to optimize these parameters (Line 4, Algorithm \ref{alg:skipgram}).
The derivatives are estimated using the back-propagation algorithm.
The learning rate $\alpha$ for SGD is initially set to 2.5\% at the beginning of the training and then decreased linearly with the number of vertices that are seen so far.

\comment{
\subsection{Complexity}
\todo{I want to show performance vs L for blogcatalog, I want to say that we converge quickly, we can reduce the size of walks in future work.}

The total length of the walks is $L = \gamma t$, then complexity of our approach is $O(dwL\log |V|)$.
Figure \ref{} shows the relation between the quality of the embeddings and length of the walks seen so far in the training.
While our model build representations that are quite useful at early stages.
The complexity of the graph structure plays a main role in how much fast we can build a good estimate of the graph topology.
As the number of possible transitions in all our walks is bounded by the number of edges $m$, we can bound our complexity to be $O(dwm\log |V|)$.
The choice of $w$ in free scale networks is bounded by the network diameter, as there is decreasing amount of information in the collocations that span the whole graph.
As $w$ is bounded by  a constant, we can see that our complexity is $O(dm\log |V|)$
}

\subsection{Parallelizability}
As shown in Figure \ref{fig:power_law} the frequency distribution of vertices in random walks of social network and words in a language both follow a power law.
This results in a long tail of infrequent vertices, therefore, the updates that affect $\Phi$ will be sparse in nature.
This allows us to use asynchronous version of stochastic gradient descent (ASGD), in the multi-worker case.
Given that our updates are sparse and we do not acquire a lock to access the model shared parameters, ASGD will  achieve an optimal rate of convergence \cite{hogwild}.
While we run experiments on one machine using multiple threads, it has been demonstrated that this technique is highly scalable, and can be used in very large scale machine learning \cite{largedeep}.
Figure \ref{fig:parallel} presents the effects of parallelizing \ouralgorithm.  It shows the speed up in processing \blogcatalog\ and \flickr\ networks is consistent as we increase the number of workers to 8 (Figure \ref{fig:parallel_speed}).
It also shows that there is no loss of predictive performance relative to the running \ouralgorithm\ serially (Figure \ref{fig:parallel_performance}).





\subsection{Algorithm Variants}
Here we discuss some variants of our proposed method, which we believe may be of interest.

\subsubsection{Streaming}
\label{variant_streaming}
One interesting variant of this method is a \emph{streaming} approach, which could be implemented without knowledge of the entire graph.
In this variant small walks from the graph are passed directly to the representation learning code, and the model is updated directly.
Some modifications to the learning process will also be necessary.
First, using a decaying learning rate will no longer be possible.  Instead, we can initialize the learning rate $\alpha$ to a small constant value.  
This will take longer to learn, but may be worth it in some applications.
Second, we cannot necessarily build a tree of parameters any more.
If the cardinality of $V$ is known (or can be bounded), we can build the Hierarchical Softmax tree for that maximum value. 
Vertices can be assigned to one of the remaining leaves when they are first seen.
If we have the ability to estimate the vertex frequency a priori, we can also still use Huffman coding to decrease frequent element access times.

\subsubsection{Non-random walks}
Some graphs are created as a by-product of agents interacting with a sequence of elements (e.g. users' navigation of pages on a website).  
When a graph is created by such a stream of \emph{non-random} walks, we can use this process to feed the modeling phase directly.
Graphs sampled in this way will not only capture information related to network structure, but also to the frequency at which paths are traversed.

In our view, this variant also encompasses language modeling. Sentences can be viewed as purposed walks through an appropriately designed language network, and 
language models like SkipGram are designed to capture this behavior.

This approach can be combined with the streaming variant (Section \ref{variant_streaming}) to train features on a continually evolving network without ever explicitly constructing the entire graph.  
Maintaining representations with this technique could enable web-scale classification without the hassles of dealing with a web-scale graph.

\begin{figure}[t!]
\centering
\begin{subfigure}[b]{0.5\columnwidth}	
\centering
\includegraphics[width=\columnwidth]{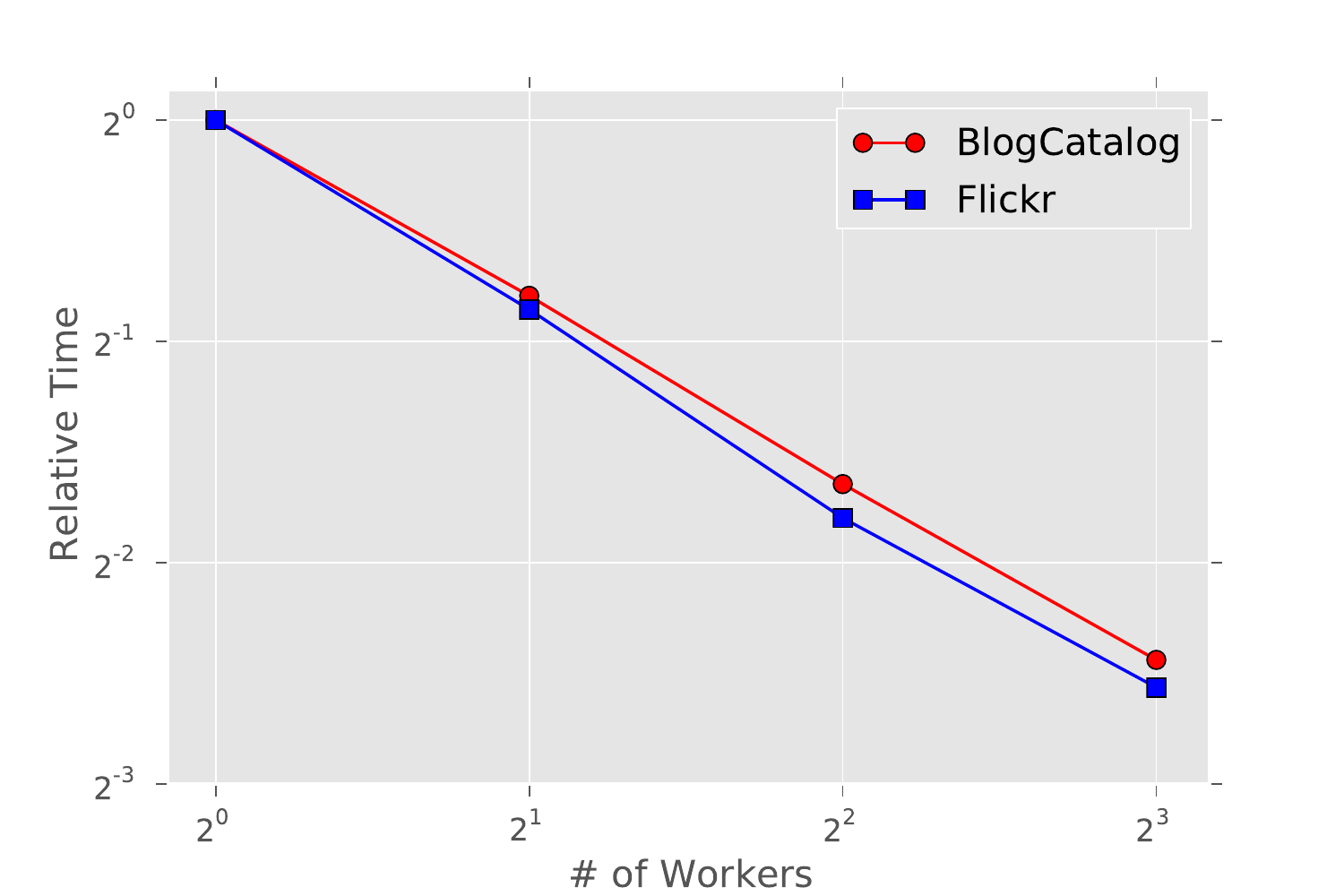}
\caption{Running Time}
\label{fig:parallel_speed}
\end{subfigure}%
\begin{subfigure}[b]{0.5\columnwidth}	
\centering
 \includegraphics[width=\columnwidth]{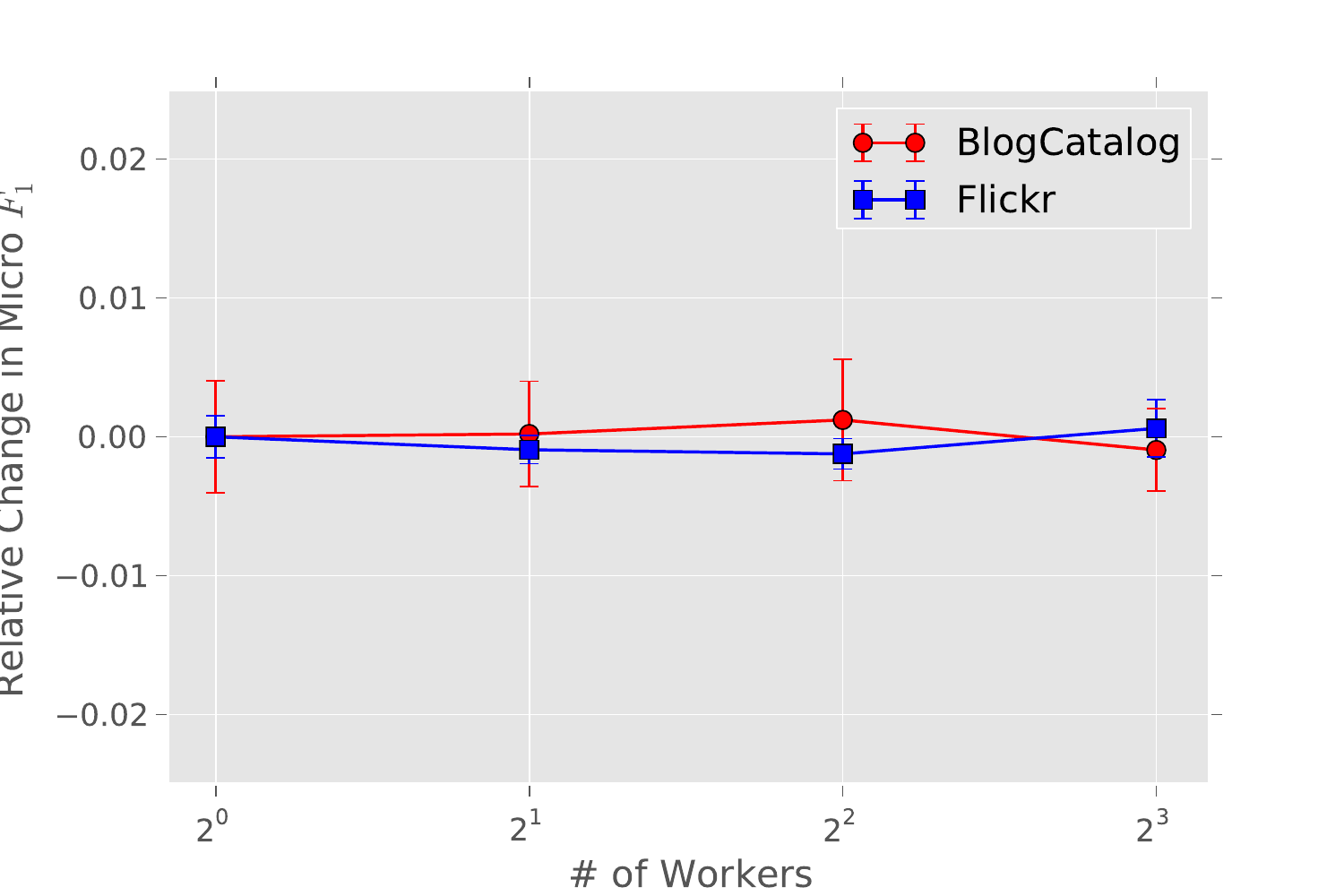}
\caption{Performance} 
\label{fig:parallel_performance}
\end{subfigure}
\caption{Effects of parallelizing \ouralgorithm}
\label{fig:parallel}
\end{figure}

\input{figures/graphs_overview}

\section{Experimental Design}
\label{sec:experimental_design}

In this section we provide an overview of the datasets and methods which we will use in our experiments.  Code and data to reproduce our results will be available at the first author's website.

\subsection{Datasets}

An overview of the graphs we consider in our experiments is given in Figure \ref{table.graph_info}.

\begin{itemize}[itemsep=1pt, topsep=5pt, partopsep=0pt]
\item \blogcatalog \cite{Tang:2009:RLV:1557019.1557109} is a network of social relationships provided by blogger authors.  The labels represent the topic categories provided by the authors.
\item \flickr \cite{Tang:2009:RLV:1557019.1557109} is a network of the contacts between users of the photo sharing website.  The labels represent the interest groups of the users such as `\emph{black and white photos}'.
\item \youtube \cite{tang2009scalable} is a social network between users of the popular video sharing website.  The labels here represent groups of viewers that enjoy common video genres (e.g. \emph{anime} and \emph{wrestling}).
\end{itemize}

\subsection{Baseline Methods}
To validate the performance of our approach we compare it against a number of baselines:

\begin{itemize}[itemsep=0pt, topsep=5pt, partopsep=0pt]
\item \socdimL \cite{Tang:2011:Leveraging}: This method generates a representation in $\mathbb{R}^d$ from the $d$-smallest eigenvectors of $\Laplacian$, the normalized graph Laplacian of $G$.  
Utilizing the eigenvectors of $\Laplacian$ implicitly assumes that graph cuts will be useful for classification.

\item \socdimB \cite{Tang:2009:RLV:1557019.1557109}: This method generates a representation in $\mathbb{R}^d$ from the top-$d$ eigenvectors of $B$, the Modularity matrix of $G$.  
The eigenvectors of $B$ encode information about modular graph partitions of $G$\cite{newman2006modularity}. Using them as features assumes that modular graph partitions will be useful for classification.

\item \socdimA \cite{tang2009scalable}: This method uses $k$-means clustering to cluster the adjacency matrix of $G$.  Its has been shown to perform comparably to the \socdimB\text{} method, with the added advantage of scaling to graphs which are too large for spectral decomposition.

\item wvRN\cite{Macskassy03asimple}: The weighted-vote Relational Neighbor is a relational classifier.
Given the neighborhood $\mathcal{N}_i$ of vertex $v_i$, wvRN estimates $\Pr(y_i|\mathcal{N}_i)$  with the (appropriately normalized) weighted mean of its neighbors (i.e $\Pr(y_i|\mathcal{N}_i) = \frac{1}{Z}\sum_{v_j \in \mathcal{N}_i}{w_{ij}\Pr(y_j \mid \mathcal{N}_j)}$).
It has shown surprisingly good performance in real networks, and has been advocated as a sensible relational classification baseline \cite{macskassy2007classification}.

\item Majority:  This na\"{\i}ve method simply chooses the most frequent labels in the training set.
\end{itemize}

\section{Experiments}
\label{sec:experiments}

\input{figures/blog_catalog_table}

\input{figures/flickr_table}

\input{figures/youtube_table}

In this section we present an experimental analysis of our method.  We thoroughly evaluate it on a number of multi-label classification tasks, and analyze its sensitivity across several parameters.

\subsection{Multi-Label Classification}
To facilitate the comparison between our method and the relevant baselines, we use the exact same datasets and experimental procedure as in \cite{Tang:2009:RLV:1557019.1557109,tang2009scalable}.
Specifically, we randomly sample a portion ($T_R$) of the labeled nodes, and use them as training data.  
The rest of the nodes are used as test.
We repeat this process 10 times, and report the average performance in terms of both Macro-$F_1$ and Micro-$F_1$. 
When possible we report the original results \cite{Tang:2009:RLV:1557019.1557109,tang2009scalable} here directly.  

For all models we use a one-vs-rest logistic regression implemented by LibLinear \cite{REF08a} for classification.
We present results for \ouralgorithm\ with ($\gamma=80$, $w=10$, $d=128$).
The results for (\socdimL, \socdimB, \socdimA) use Tang and Liu's preferred dimensionality, $d=500$.

\subsubsection{BlogCatalog}
In this experiment we increase the training ratio ($T_R$) on the \blogcatalog\text{} network from 10\% to 90\%.
Our results are presented in Table \ref{tbl:blogcatalog}.  Numbers in bold represent the highest performance in each column.  

\ouralgorithm\ performs consistently better than \socdimA, \socdimB, and \wvrn.  
In fact, when trained with only 20\% of the nodes labeled, \ouralgorithm\ performs better than these approaches when they are given 90\% of the data.
The performance of \socdimL\ proves much more competitive, but \ouralgorithm\ still outperforms when labeled data is sparse on both Macro-$F_1$ ($T_R\leq20\%$) and Micro-$F_1$ ($T_R\leq60$\%).  

This strong performance when only small fractions of the graph are labeled is a core strength of our approach.  
In the following experiments, we investigate the performance of our representations on even more sparsely labeled graphs.

\input{figures/stability.tex}

\subsubsection{Flickr}
In this experiment we vary the training ratio ($T_R$) on the \flickr\ network from 1\% to 10\%.  
This corresponds to having approximately 800 to 8,000 nodes labeled for classification in the entire network. 
Table \ref{tbl:flickr} presents our results, which are consistent with the previous experiment.  \ouralgorithm\ outperforms all baselines by at least 3\% with respect to Micro-$F_1$. 
Additionally, its Micro-$F_1$ performance when only 3\% of the graph is labeled beats all other methods even when they have been given 10\% of the data.
In other words, \ouralgorithm\ can outperform the baselines with 60\% less training data. 
It also performs quite well in Macro-$F_1$, initially performing close to \socdimL, but distancing itself to a 1\% improvement.

\subsubsection{YouTube}
The \youtube\ network is considerably larger than the previous ones we have experimented on, and its size prevents two of our baseline methods (\socdimL\text{} and \socdimB) from running on it.
It is much closer to a real world graph than those we have previously considered.

The results of varying the training ratio ($T_R$) from 1\% to 10\% are presented in Table \ref{tbl:youtube}.
They show that \ouralgorithm\ significantly outperforms the scalable baseline for creating graph representations, \socdimA.  When 1\% of the labeled nodes are used for test, the Micro-$F_1$ improves by 14\%.
The Macro-$F_1$ shows a corresponding 10\% increase.
This lead narrows as the training data increases, but \ouralgorithm\ ends with a 3\% lead in Micro-$F_1$, and an impressive 5\% improvement in Macro-$F_1$.

This experiment showcases the performance benefits that can occur from using social representation learning for multi-label classification.  
\ouralgorithm, can scale to large graphs, and performs exceedingly well in such a sparsely labeled environment.

\subsection{Parameter Sensitivity}
In order to evaluate how changes to the parameterization of \ouralgorithm\ effect its performance on classification tasks, we conducted experiments on two multi-label classifications tasks (\flickr, and \blogcatalog).
For this test, we have fixed the window size and the walk length to sensible values $(w=10,t=40)$ which should emphasize local structure. 
We then vary the number of latent dimensions ($d$), the number of walks started per vertex ($\mathcal{\gamma}$), and the amount of training data available ($T_R$) to determine their impact on the network classification performance.

\subsubsection{Effect of Dimensionality}
Figure \ref{fig:stability_dimensions} shows the effects of increasing the number of latent dimensions available to our model.

Figures \ref{fig:stability_flickr-dims_vs_training} and \ref{fig:stability_blogcatalog-dims_vs_training} examine the effects of varying the dimensionality and training rate.
The performance is quite consistent between both \flickr\ and \blogcatalog\, and show that the optimal dimensionality for a model is dependent on the number of training examples.  (Note that 1\% of \flickr\ has approximately as many labeled examples as 10\% of \blogcatalog).

Figures \ref{fig:stability_flickr-dims_vs_passes} and \ref{fig:stability_blogcatalog-dims_vs_training} examine the effects of varying the dimensionality and number of walks per vertex.
The relative performance between dimensions is relatively stable across different values of $\gamma$.
These charts have two interesting observations.  The first is that there is most of the benefit is accomplished by starting $\gamma = 30$ walks per node in both graphs.
The second is that the relative difference between different values of $\gamma$ is quite consistent between the two graphs.
\flickr\ has an order of magnitude more edges than \blogcatalog, and we find this behavior interesting.

These experiments show that our method can make useful models of various sizes.  They also show that the performance of the model depends on the number of random walks it has seen, and the appropriate dimensionality of the model depends on the training examples available.

\subsubsection{Effect of sampling frequency}
Figure \ref{fig:stability_passes} shows the effects of increasing $\gamma$, the number of random walks that we start from each vertex.  

The results are very consistent for different dimensions (Fig.\ \ref{fig:stability_flickr-passes_vs_dims}, Fig.\ \ref{fig:stability_blogcatalog-passes_vs_dims}) and the amount of training data (Fig.\ \ref{fig:stability_flickr-passes_vs_training}, Fig.\ \ref{fig:stability_blogcatalog-passes_vs_training}).
Initially, increasing $\gamma$ has a big effect in the results, but this effect quickly slows ($\gamma > 10$).
These results demonstrate that we are able to learn meaningful latent representations for vertices after only a small number of random walks.

\comment{
\section{Discussion}
\label{sec:discussion}

\subsection{Why does it work?}

\subsubsection{H1. Random walk is a better way to de-sparsify than PCA on $(A, L, B)$}
Idea:  Machine learning is easier on dense things, and some methods make better dense things

\subsubsection{H2. Decoupling feature space learning from label space avoids cascading errors.}

\subsubsection{H3. Short random walks are enough (partial topology)}

\subsubsection{H4. Plotting these things in 2-D will reveal differences on Cora}

\subsection{Limitations}
\todo{Improve this}
\subsubsection{Homophily Assumption}
Using random walks to extract information from graphs places limits on the type of information that we can hope to extract from the network.
Specifically, random walk methods have a strong homophily assumption, expecting that random walk distance is correlated in some way with the output space.
While this assumption holds in many real world in many real-world graphs, it might not be a good assumption in some (e.g. in bipartite networks \cite{Gallagher:2008:UGE:1401890.1401925}).
This emphasis on homophily also means that random walk methods can not effectively capture heterophily as some relational classification approaches can \cite{neville2000iterative}.

\subsubsection{Scale}
The largest correlations we can hope to model are limited by the length of the window.
}

\section{Related Work}
\label{sec:related}
The main differences between our proposed method and previous work can be summarized as follows: 
\begin{enumerate} [itemsep=0pt, topsep=5pt, partopsep=0pt]
\item We \emph{learn} our latent social representations, instead of computing statistics related to centrality \cite{gallagher2010leveraging} or partitioning \cite{Tang:2011:Leveraging}.
\item We do not attempt to extend the classification procedure itself (through collective inference \cite{sen2008collective} or graph kernels \cite{kondor2002diffusion}).
\item We propose a scalable online method which uses only local information.  Most methods require global information and are offline \cite{Tang:2009:RLV:1557019.1557109,tang2009scalable,Tang:2011:Leveraging,Henderson:2011:YKG:2020408.2020512}.
\item We apply unsupervised representation learning to graphs.
\end{enumerate}
In this section we discuss related work in network classification and unsupervised feature learning.

\subsection{Relational Learning}
Relational classification (or \emph{collective classification}) methods \cite{Macskassy03asimple,neville2000iterative,Pearl:1988:PRI:534975,geman1984stochastic} use links between data items as part of the classification process. 
Exact inference in the collective classification problem is NP-hard, and solutions have focused on the use of approximate inference algorithm which may not be guaranteed to converge \cite{sen2008collective}.

The most relevant relational classification algorithms to our work incorporate community information by learning clusters \cite{Neville:2005:LRA:1090193.1090201},  by adding edges between nearby nodes \cite{Gallagher:2008:UGE:1401890.1401925}, by using PageRank \cite{cohenASOM}, or by extending relational classification to take additional features into account \cite{wang2013multi}.
Our work takes a substantially different approach.  
Instead of a new approximation inference algorithm, we propose a procedure which learns representations of network structure which can then be used by existing inference procedure (including iterative ones).  

A number of techniques for generating features from graphs have also been proposed \cite{gallagher2010leveraging,Tang:2009:RLV:1557019.1557109,tang2009scalable,Tang:2011:Leveraging,Henderson:2011:YKG:2020408.2020512}.
In contrast to these methods, we frame the feature creation procedure as a representation learning problem.

Graph Kernels \cite{vishwanathan2010graph} have been proposed as a way to use relational data as part of the classification process, but are quite slow unless approximated \cite{kang2012fast}.  
Our approach is complementary;  instead of encoding the structure as part of a kernel function, we learn a representation which allows them to be used directly as features for any classification method.

\subsection{Unsupervised Feature Learning}
Distributed representations have been proposed to model structural relationship between concepts \cite{distributed}.
These representations are trained by the back-propagation and gradient descent.
Computational costs and numerical instability led to these techniques to be abandoned for almost a decade.
Recently, distributed computing allowed for larger models to be trained \cite{nnlm}, and the growth of data for unsupervised learning algorithms to emerge \cite{erhanhelps}.
Distributed representations usually are trained through neural networks, these networks have made advancements in diverse fields such as computer vision \cite{vision1}, speech recognition \cite{speech1}, and natural language processing \cite{senna1}.

\section{Conclusions}
\label{sec:conclusion}
We propose \ouralgorithm, a novel approach for learning latent social representations of vertices. 
Using local information from truncated random walks as input, our method learns a representation which encodes structural regularities.
Experiments on a variety of different graphs illustrate the effectiveness of our approach on challenging multi-label classification tasks.

As an online algorithm, \ouralgorithm\textsc{} is also scalable. Our results show that we can create meaningful representations for graphs too large to run spectral methods on.
On such large graphs, our method significantly outperforms other methods designed to operate for sparsity.
We also show that our approach is parallelizable, allowing workers to update different parts of the model concurrently.

In addition to being effective and scalable, our approach is also an appealing generalization of language modeling. 
This connection is mutually beneficial.  
Advances in language modeling may continue to generate improved latent representations for networks.
In our view, language modeling is actually sampling from an unobservable language graph.  We believe that insights obtained from modeling observable graphs may in turn yield improvements to modeling unobservable ones.

Our future work in the area will focus on investigating this duality further, using our results to improve language modeling, and strengthening the theoretical justifications of the method.

\bibliography{paper}
\bibliographystyle{abbrv}

\end{document}


\maketitle

\section*{Appendix}

\subsection{Streaming}
You don't even need a graph.  You don't even need a row.  You only need a \emph{sample} from a row's distribution!

\subsection{Structure Preserving Properties}

\subsection{Scalability}

%% file: figures/graphs_overview.tex
\begin{table}[t!]
\begin{center}
  \begin{tabular}{ c | c | c | c }
	Name & \blogcatalog & \flickr & \youtube \\
    \hline
	$|V|$ & 10,312 & 80,513 & 1,138,499 \\
	$|E|$ & 333,983 & 5,899,882 & 2,990,443 \\
	$|\mathcal{Y}|$ & 39 & 195 & 47 \\
	Labels & Interests & Groups & Groups \\

  \end{tabular}
  \caption{Graphs used in our experiments.}
  \label{table.graph_info}  
\end{center}
\end{table}

%% file: figures/blog_catalog_table.tex
\begin{table*}[p]
\begin{center}

\begin{tabular}{l|l|r|r|r|r|r|r|r|r|r}
 & \% Labeled Nodes & 10\% & 20\% & 30\% & 40\% & 50\% & 60\% & 70\% & 80\% & 90\% \\ \hline
  & & & & & & & & & & \\ 
  \rowcolor{high}
 & \ouralgorithm & \textbf{36.00} & \textbf{38.20} & \textbf{39.60} & \textbf{40.30} & 
 \textbf{41.00} & \textbf{41.30} & 41.50 & 41.50 & 42.00 \\
 & \socdimL & 31.06 & 34.95 & 37.27 & 38.93 & 39.97 & 40.99 & \textbf{41.66} & \textbf{42.42} & \textbf{42.62} \\
 & \socdimA & 27.94 & 30.76 & 31.85 & 32.99 & 34.12 & 35.00 & 34.63 & 35.99 & 36.29 \\
Micro-F1(\%) & \socdimB & 27.35 & 30.74 & 31.77 & 32.97 & 34.09 & 36.13 & 36.08 & 37.23 & 38.18 \\
 & wvRN & 19.51 & 24.34 & 25.62 & 28.82 & 30.37 & 31.81 & 32.19 & 33.33 & 34.28 \\
 & Majority & 16.51 & 16.66 & 16.61 & 16.70 & 16.91 & 16.99 & 16.92 & 16.49 & 17.26 \\
 & & & & & & & & & & \\ \hline
 & & & & & & & & & & \\
 \rowcolor{high} 
 & \ouralgorithm & \textbf{21.30} & \textbf{23.80} & 25.30 & 26.30 & 27.30 & 27.60 & 27.90 & 28.20 & 28.90 \\ 
 & \socdimL & 19.14 & 23.57 & \textbf{25.97} & \textbf{27.46} & \textbf{28.31} & \textbf{29.46} & \textbf{30.13} & \textbf{31.38} & \textbf{31.78} \\
 & \socdimA & 16.16 & 19.16 & 20.48 & 22.00 & 23.00 & 23.64 & 23.82 & 24.61 & 24.92 \\
Macro-F1(\%) & \socdimB & 17.36 & 20.00 & 20.80 & 21.85 & 22.65 & 23.41 & 23.89 & 24.20 & 24.97 \\
 & wvRN & 6.25 & 10.13 & 11.64 & 14.24 & 15.86 & 17.18 & 17.98 & 18.86 & 19.57 \\ 
 & Majority & 2.52 & 2.55 & 2.52 & 2.58 & 2.58 & 2.63 & 2.61 & 2.48 & 2.62 \\
\end{tabular}
\end{center}
\caption{Multi-label classification results in \blogcatalog}
\label{tbl:blogcatalog}
\end{table*}

%% file: figures/flickr_table.tex
\begin{table*}[p]
\begin{center}
\begin{tabular}{l|l|r|r|r|r|r|r|r|r|r|r}
 & \% Labeled Nodes & 1\% & 2\% & 3\% & 4\% & 5\% & 6\% & 7\% & 8\% & 9\% & 10\% \\ \hline
 & & & & & & & & & & & \\
 \rowcolor{high} 
 & \ouralgorithm & \textbf{32.4} & \textbf{34.6} & \textbf{35.9} & \textbf{36.7} & \textbf{37.2} & \textbf{37.7} & \textbf{38.1} & \textbf{38.3} & \textbf{38.5} & \textbf{38.7} \\
 & \socdimL & 27.43 & 30.11 & 31.63 & 32.69 & 33.31 & 33.95 & 34.46 & 34.81 & 35.14 & 35.41 \\
Micro-F1(\%) & \socdimA & 25.75 & 28.53 & 29.14 & 30.31 & 30.85 & 31.53 & 31.75 & 31.76 & 32.19 & 32.84 \\
 & \socdimB & 22.75 & 25.29 & 27.3 & 27.6 & 28.05 & 29.33 & 29.43 & 28.89 & 29.17 & 29.2 \\
 & wvRN & 17.7 & 14.43 & 15.72 & 20.97 & 19.83 & 19.42 & 19.22 & 21.25 & 22.51 & 22.73 \\
 & Majority & 16.34 & 16.31 & 16.34 & 16.46 & 16.65 & 16.44 & 16.38 & 16.62 & 16.67 & 16.71 \\
 & & & & & & & & & & & \\ \hline
 & & & & & & & & & & & \\
 \rowcolor{high}
 & \ouralgorithm & \textbf{14.0} & 17.3 & \textbf{19.6} & \textbf{21.1} & \textbf{22.1} & \textbf{22.9} & \textbf{23.6} & \textbf{24.1} & \textbf{24.6} & \textbf{25.0} \\
 & \socdimL & 13.84 & \textbf{17.49} & 19.44 & 20.75 & 21.60 & 22.36 & 23.01 & 23.36 & 23.82 & 24.05 \\
Macro-F1(\%) & \socdimA & 10.52 & 14.10 & 15.91 & 16.72 & 18.01 & 18.54 & 19.54 & 20.18 & 20.78 & 20.85 \\
 & \socdimB & 10.21 & 13.37 & 15.24 & 15.11 & 16.14 & 16.64 & 17.02 & 17.1 & 17.14 & 17.12 \\
 & wvRN & 1.53 & 2.46 & 2.91 & 3.47 & 4.95 & 5.56 & 5.82 & 6.59 & 8.00 & 7.26 \\
 & Majority & 0.45 & 0.44 & 0.45 & 0.46 & 0.47 & 0.44 & 0.45 & 0.47 & 0.47 & 0.47 \\
\end{tabular}
\end{center}
\caption{Multi-label classification results in \flickr}
\label{tbl:flickr}
\end{table*}

%% file: figures/youtube_table.tex
\begin{table*}[p]
\begin{center}
\begin{tabular}{l|l|r|r|r|r|r|r|r|r|r|r}
 & \% Labeled Nodes & 1\% & 2\% & 3\% & 4\% & 5\% & 6\% & 7\% & 8\% & 9\% & 10\% \\ \hline
 & & & & & & & & & & \\
 \rowcolor{high}
 & \ouralgorithm & \textbf{37.95} & \textbf{39.28} & \textbf{40.08} & \textbf{40.78} & \textbf{41.32} & \textbf{41.72} & \textbf{42.12} & \textbf{42.48} & \textbf{42.78} & \textbf{43.05} \\ 
 & \socdimL & --- & --- & --- & --- & --- & --- & --- & --- & ---  & --- \\ 
Micro-F1(\%) & \socdimA  & 23.90 & 31.68 & 35.53 & 36.76 & 37.81 & 38.63 & 38.94 & 39.46 & 39.92 & 40.07 \\
 & \socdimB & --- & --- & --- & --- & --- & --- & --- & --- & --- & --- \\ 
 & wvRN & 26.79 & 29.18 & 33.1 & 32.88 & 35.76 & 37.38 & 38.21 & 37.75 & 38.68 & 39.42 \\ 
 & Majority & 24.90 & 24.84 & 25.25 & 25.23 & 25.22 & 25.33 & 25.31 & 25.34 & 25.38 & 25.38 \\ 
  & & & & & & & & & & \\ \hline
 & & & & & & & & & & \\
 \rowcolor{high}
 & \ouralgorithm & \textbf{29.22} & \textbf{31.83} & \textbf{33.06} & \textbf{33.90} & \textbf{34.35} & \textbf{34.66} & \textbf{34.96} & \textbf{35.22} & \textbf{35.42} & \textbf{35.67} \\
 & \socdimL & --- & --- & --- & --- & --- & --- & --- & --- & --- & --- \\ 
Macro-F1(\%) & \socdimA  & 19.48 & 25.01 & 28.15 & 29.17 & 29.82 & 30.65 & 30.75 & 31.23 & 31.45 & 31.54 \\ 
 & \socdimB & --- & --- & --- & --- & --- & --- & --- & --- & --- & --- \\ 
 & wvRN & 13.15 & 15.78 & 19.66 & 20.9 & 23.31 & 25.43 & 27.08 & 26.48 & 28.33 & 28.89 \\
 & Majority & 6.12 & 5.86 & 6.21 & 6.1 & 6.07 & 6.19 & 6.17 & 6.16 & 6.18 & 6.19 \\
\end{tabular}
\end{center}
\caption{Multi-label classification results in \youtube}
\label{tbl:youtube}
\end{table*}

%% file: figures/stability.tex
\begin{figure*}[t!]
	\centering	
		\begin{subfigure}[b]{\columnwidth}	
			\begin{subfigure}[b]{\columnwidth}
    		    \begin{subfigure}[b]{0.5\columnwidth}    
    		    		\centering
    	            \includegraphics[width=\columnwidth]{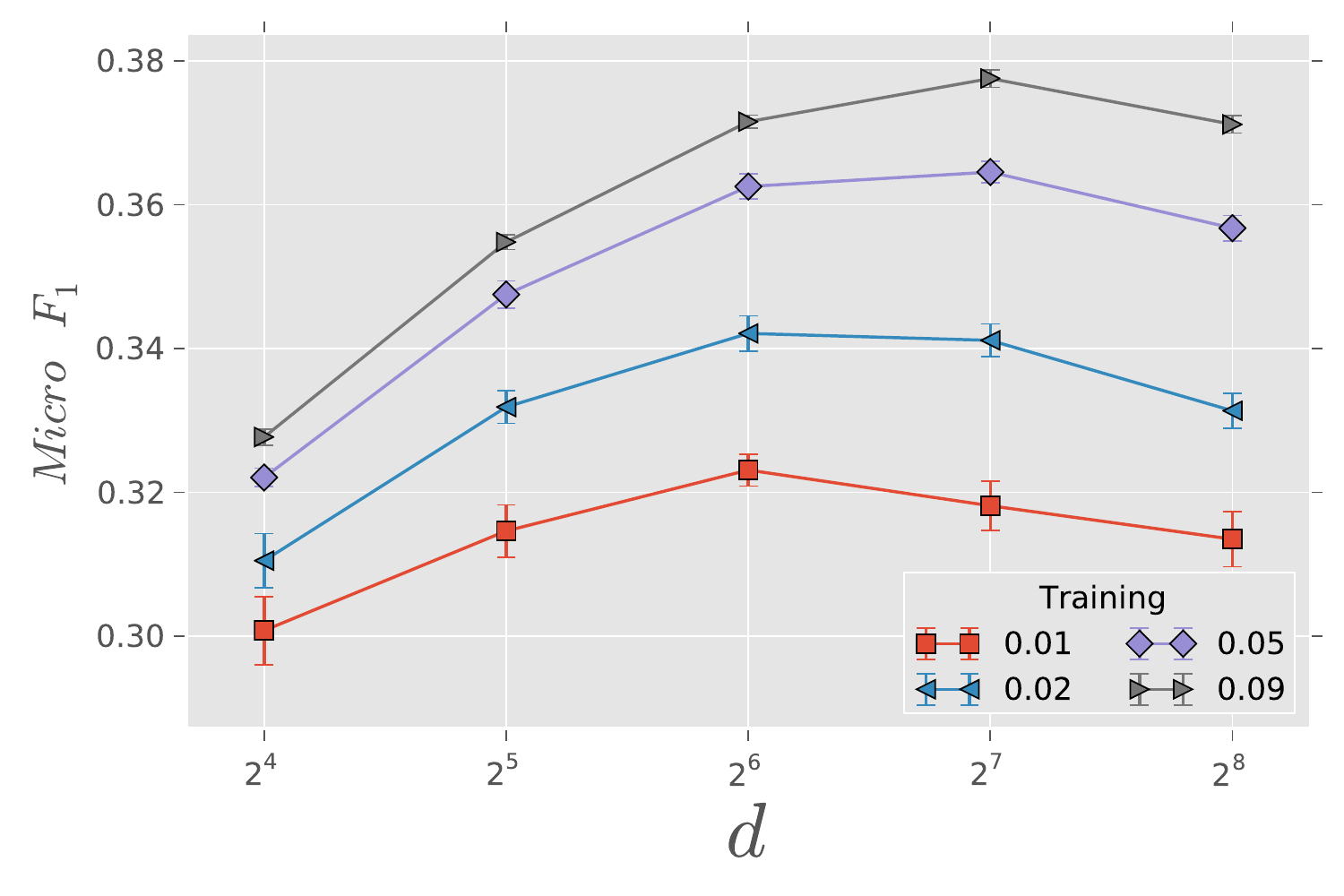}%
			        \renewcommand\thesubfigure{\alph{subfigure}1}%
	                \caption{\flickr, $\gamma=30$}    	            
    	            \label{fig:stability_flickr-dims_vs_training}
	    	    \end{subfigure}%
	        \begin{subfigure}[b]{0.5\columnwidth}
    		    		\centering	        
	                \includegraphics[width=\columnwidth]{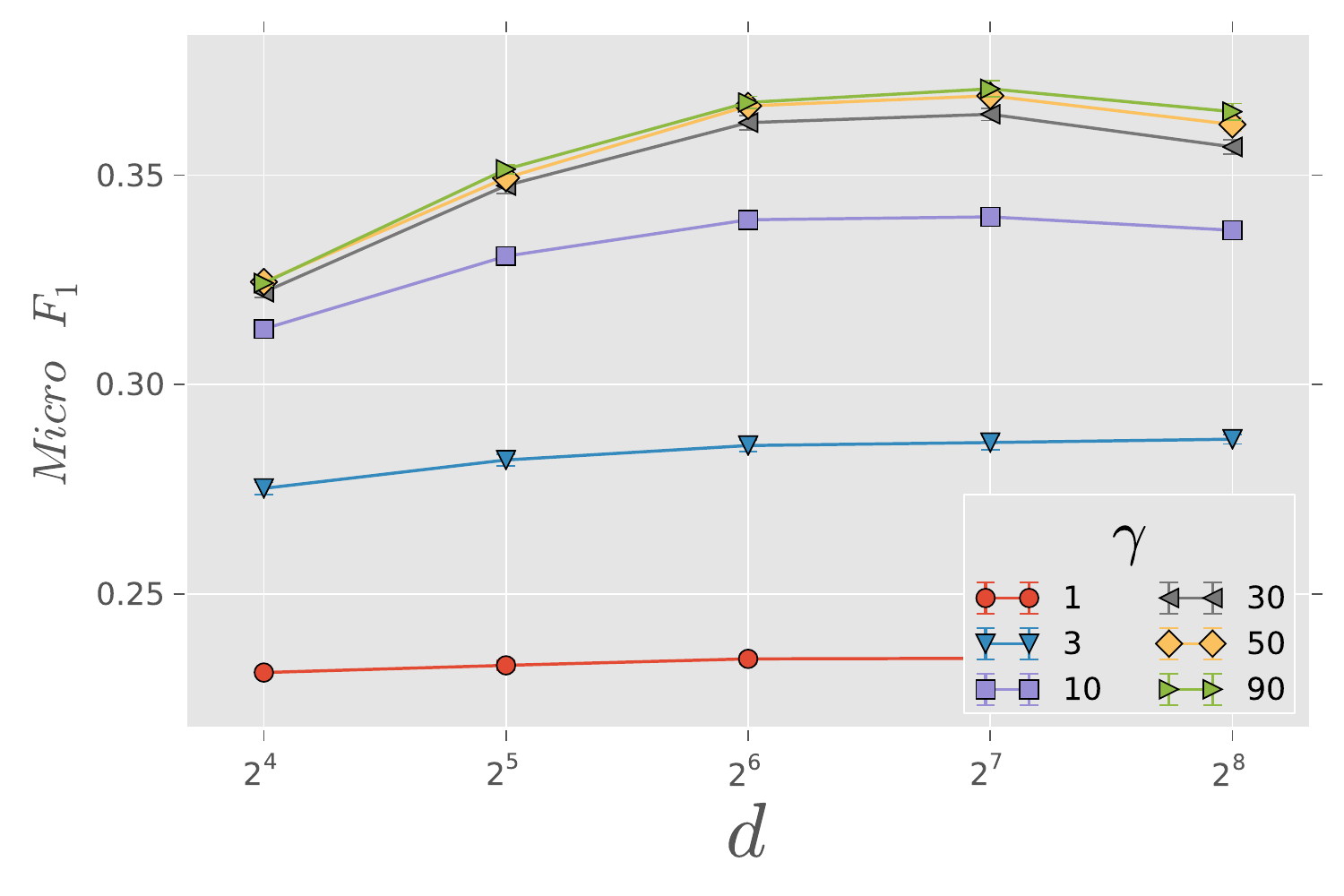}%
				    \addtocounter{subfigure}{-1}%
			        \renewcommand\thesubfigure{\alph{subfigure}2}%
	                \caption{\flickr, $T_R=0.05$}
	                \label{fig:stability_flickr-dims_vs_passes}
	        \end{subfigure}%
        \end{subfigure}%
		\hfill
		\begin{subfigure}[b]{\columnwidth} %
        \begin{subfigure}{0.5\columnwidth} %
                \includegraphics[width=\columnwidth]{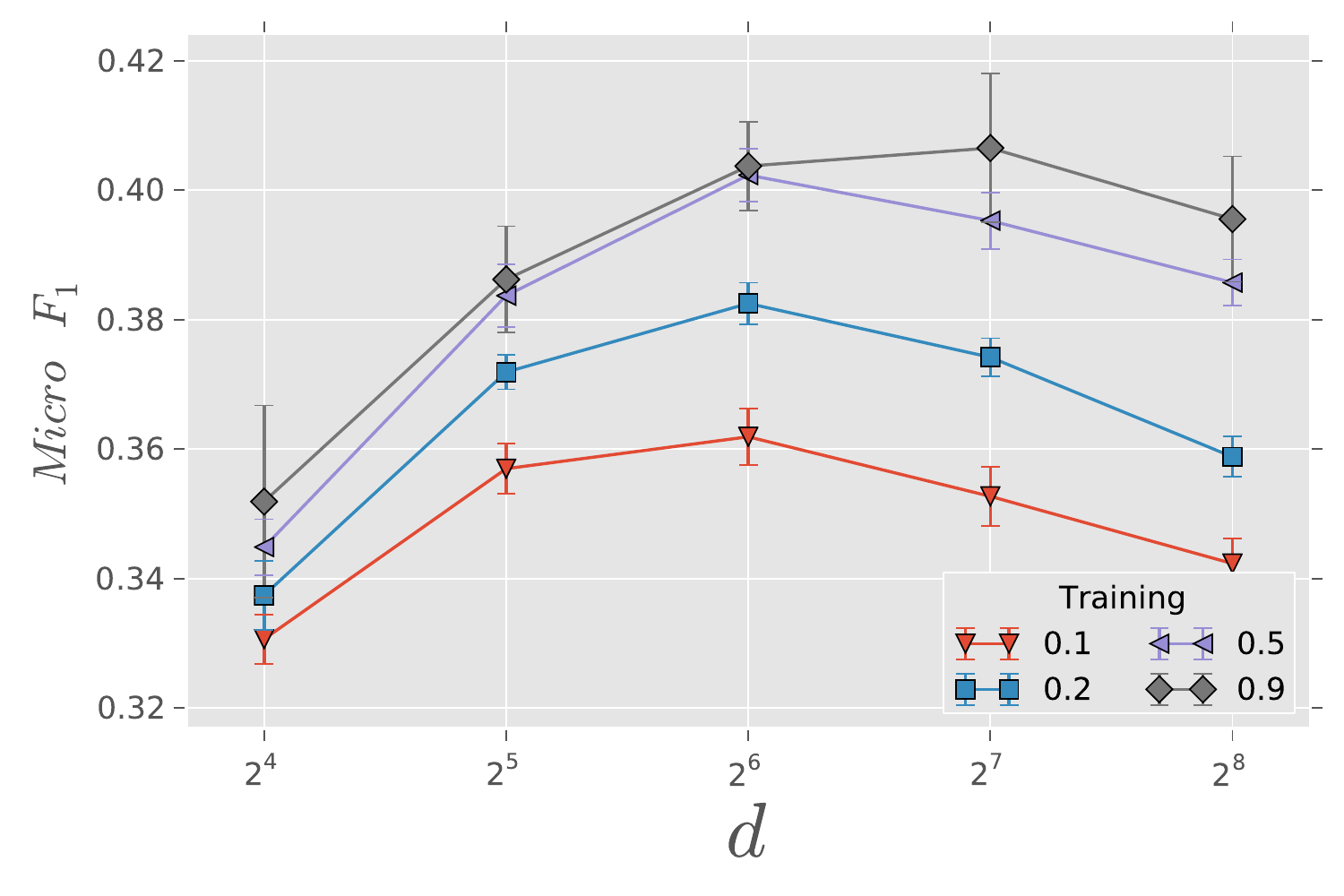}%
				    \addtocounter{subfigure}{-1}%
			        \renewcommand\thesubfigure{\alph{subfigure}3}%
	                \caption{\blogcatalog, $\gamma=30$}                
        \label{fig:stability_blogcatalog-dims_vs_training}	                
        \end{subfigure}%
        \begin{subfigure}{0.5\columnwidth}%
                \includegraphics[width=\columnwidth]{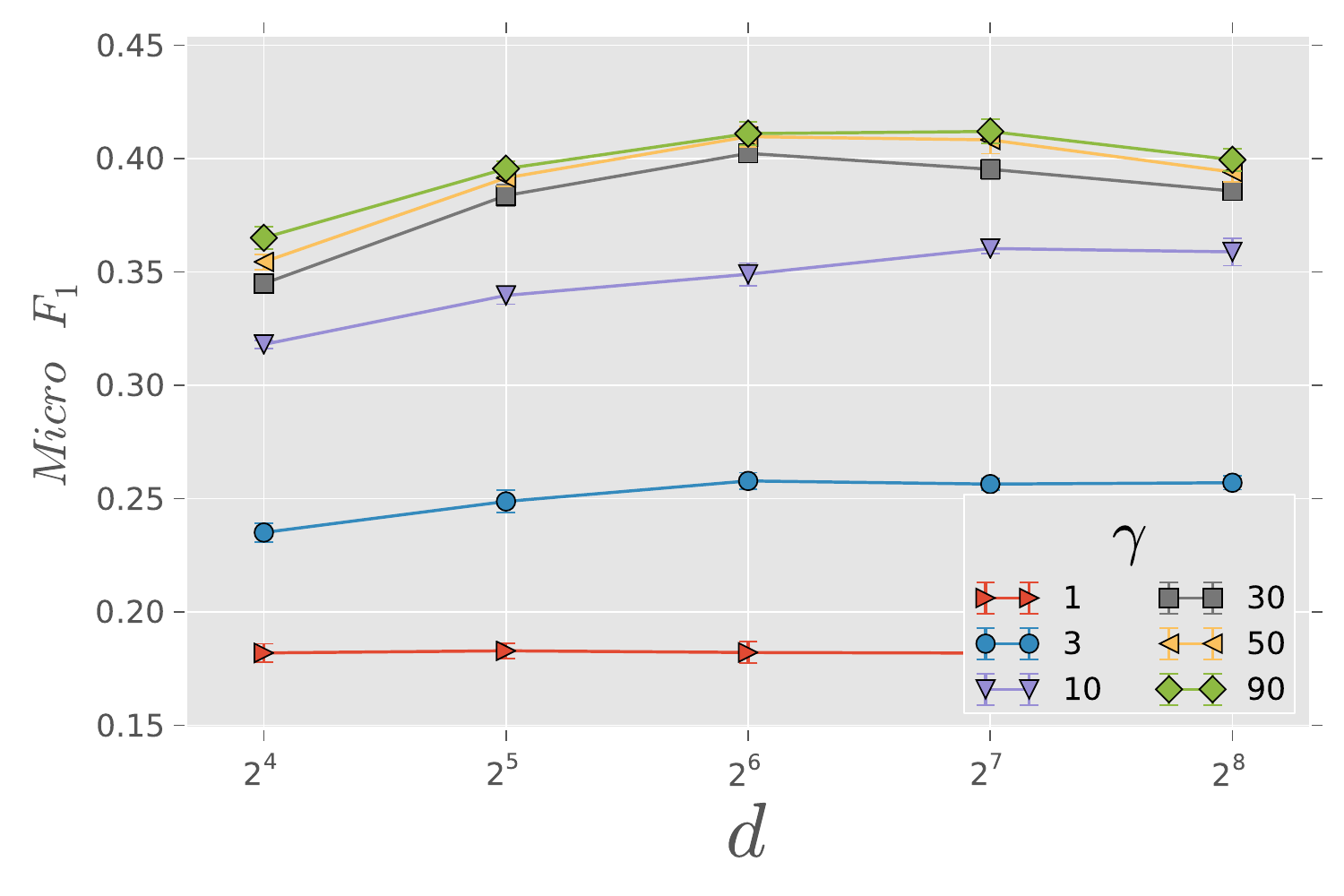}%
				    \addtocounter{subfigure}{-1}%
			        \renewcommand\thesubfigure{\alph{subfigure}4} %
	                \caption{\blogcatalog, $T_R=0.5$} %
		        \label{fig:stability_blogcatalog-dims_vs_passes}   
        \end{subfigure}        
        \label{fig:stability_blogcatalog-dims}
        \end{subfigure}%
        \addtocounter{subfigure}{-1}        
		\caption{Stability over dimensions, $d$}
		\label{fig:stability_dimensions}
		\end{subfigure}%
		\hfill
		\begin{subfigure}[b]{\columnwidth}			
		\begin{subfigure}[b]{\columnwidth}
    		    \begin{subfigure}[b]{0.5\columnwidth}
    	            \includegraphics[width=\columnwidth]{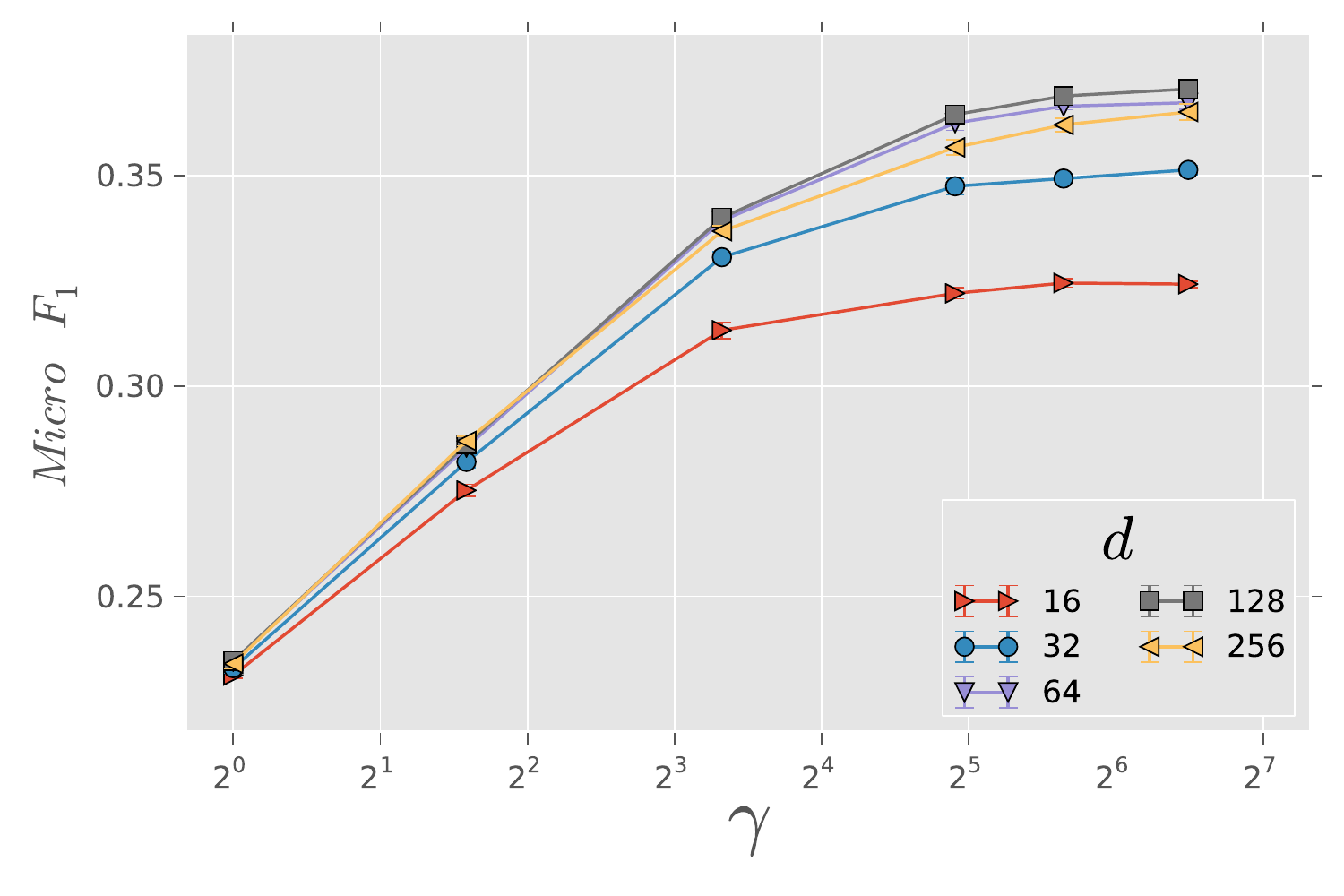}%
			        \renewcommand\thesubfigure{\alph{subfigure}1}%
	                \caption{\flickr, $T_R=0.05$}    	            
    	            \label{fig:stability_flickr-passes_vs_dims}    	            
	    	    \end{subfigure}%
	        \begin{subfigure}[b]{0.5\columnwidth}
	                \includegraphics[width=\columnwidth]{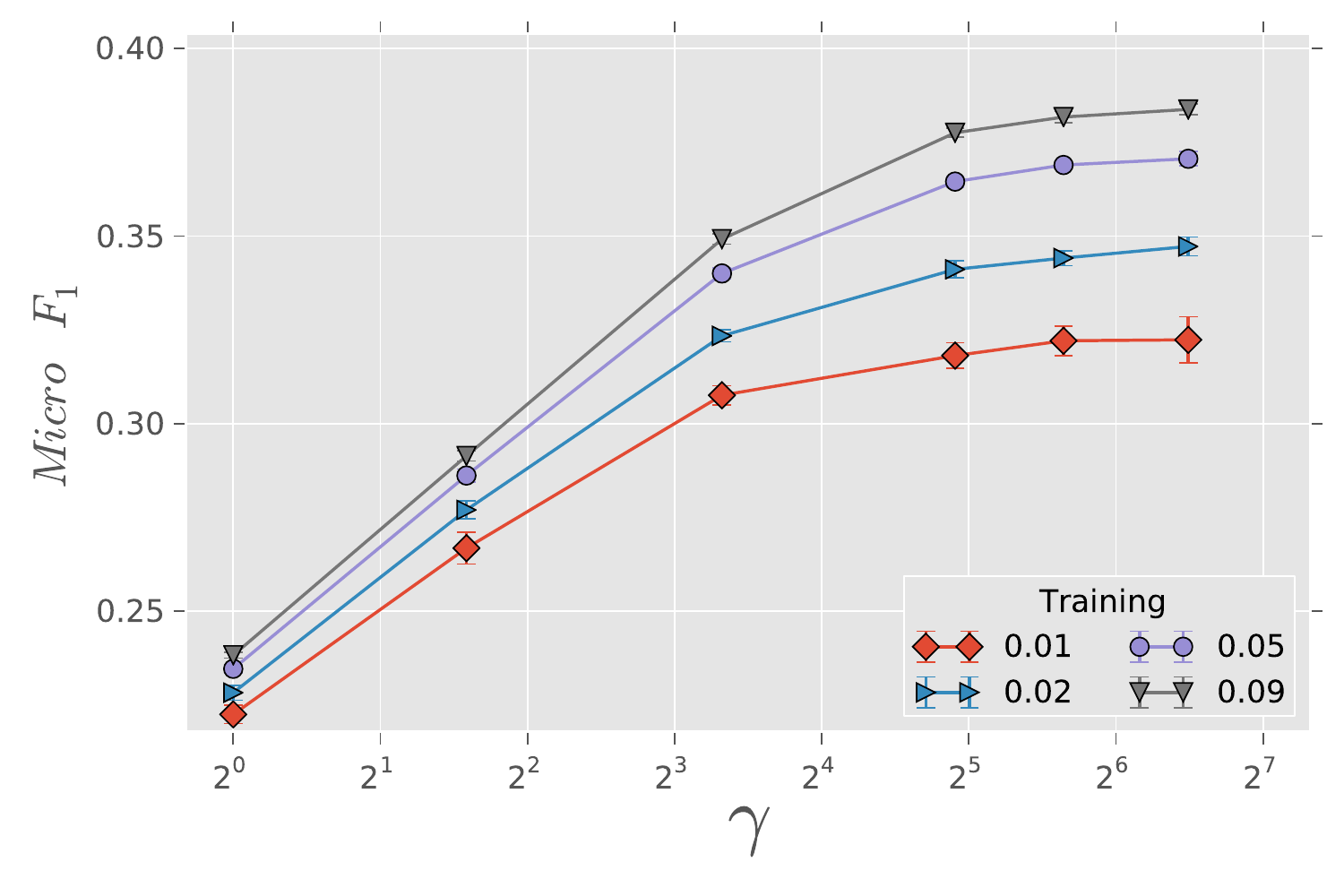}%
				    \addtocounter{subfigure}{-1}%
			        \renewcommand\thesubfigure{\alph{subfigure}2}%
	                \caption{\flickr, $d=128$}
	                \label{fig:stability_flickr-passes_vs_training}	                
	        \end{subfigure}%
        \end{subfigure}%
		\hfill
		\begin{subfigure}[b]{\columnwidth}        
        \begin{subfigure}{0.5\columnwidth}
                \includegraphics[width=\columnwidth]{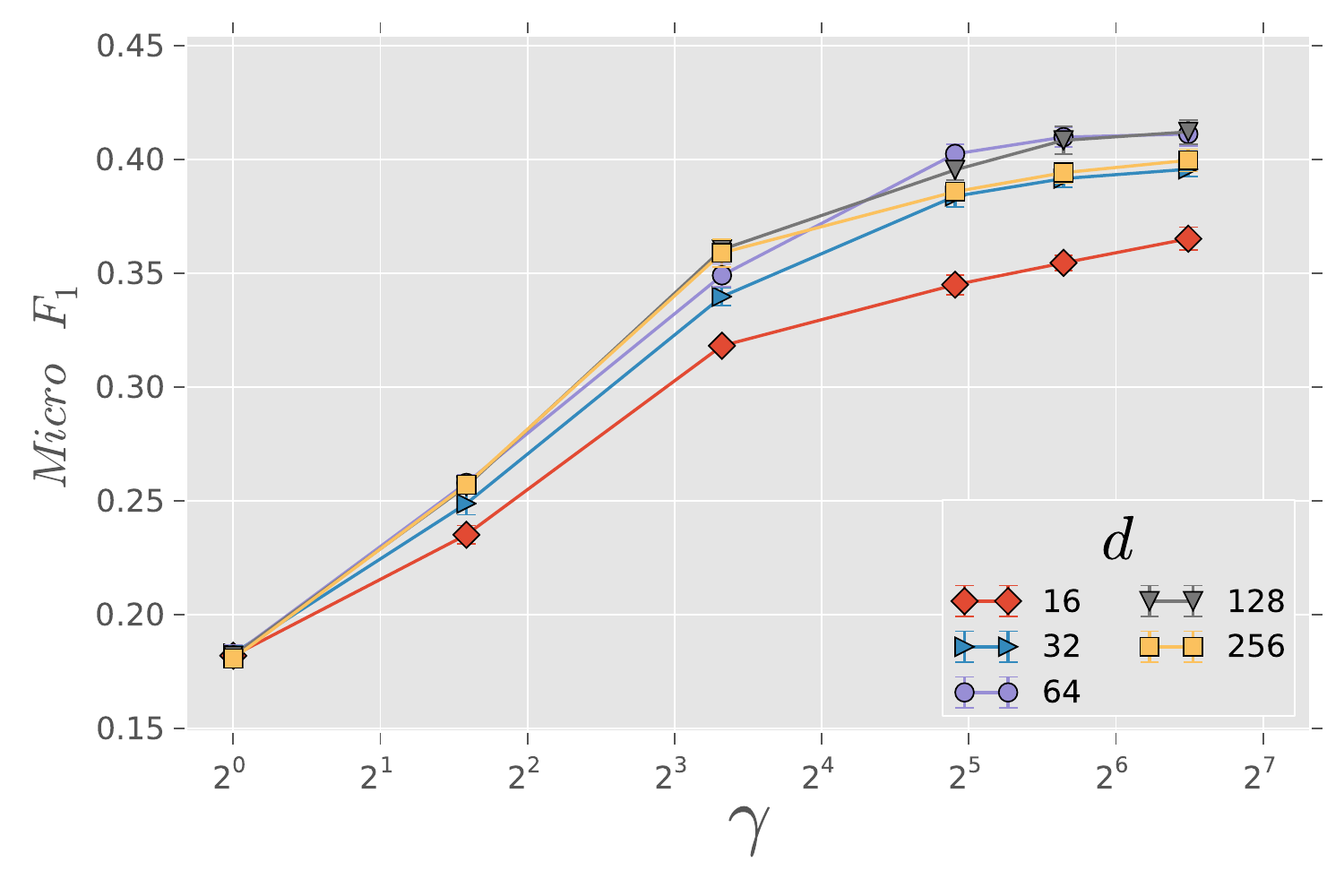}%
			    \addtocounter{subfigure}{-1}%
		       \renewcommand\thesubfigure{\alph{subfigure}3}%
               \caption{\blogcatalog,  $T_R=0.5$}                
			   \label{fig:stability_blogcatalog-passes_vs_dims}	                   
        \end{subfigure}%
        \begin{subfigure}{0.5\columnwidth}
                \includegraphics[width=\columnwidth]{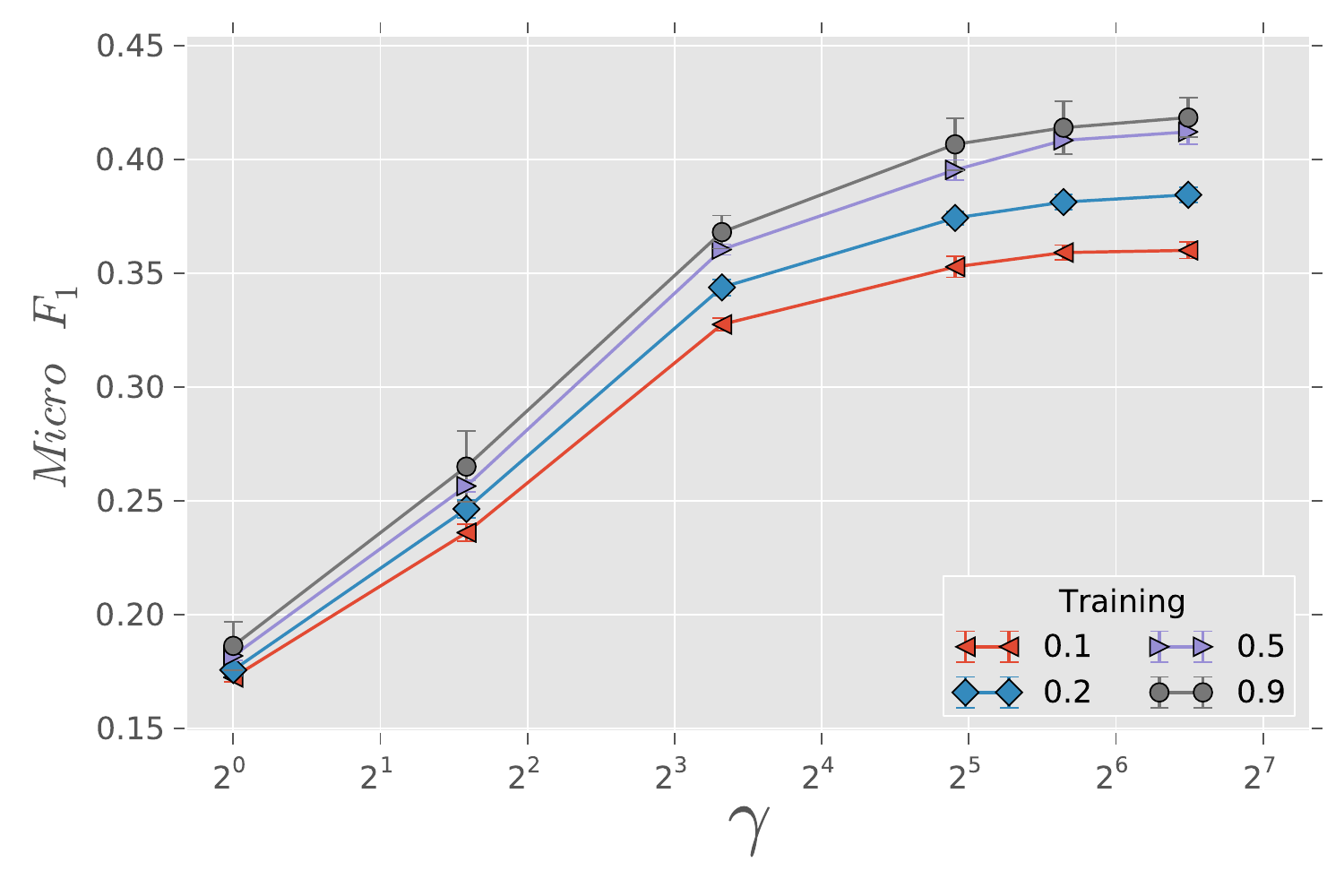}%
			    \addtocounter{subfigure}{-1}%
		        \renewcommand\thesubfigure{\alph{subfigure}4} %
                \caption{\blogcatalog, $d=128$} %
		        \label{fig:stability_blogcatalog-passes_vs_training}                  
        \end{subfigure}  
        \addtocounter{subfigure}{-1}
        \end{subfigure}%
        \addtocounter{subfigure}{-1}                                 
		\caption{Stability over number of walks, $\gamma$}
		\label{fig:stability_passes}
		\end{subfigure}%
\caption{Parameter Sensitivity Study}
\label{fig:stability}
\end{figure*}

%% file: paper.bbl
\begin{thebibliography}{10}

\bibitem{andersen2006local}
R.~Andersen, F.~Chung, and K.~Lang.
\newblock Local graph partitioning using pagerank vectors.
\newblock In {\em Foundations of Computer Science, 2006. FOCS'06. 47th Annual
  IEEE Symposium on}, pages 475--486. IEEE, 2006.

\bibitem{deepfuture}
Y.~Bengio, A.~Courville, and P.~Vincent.
\newblock Representation learning: A review and new perspectives.
\newblock 2013.

\bibitem{nnlm}
Y.~Bengio, R.~Ducharme, and P.~Vincent.
\newblock A neural probabilistic language model.
\newblock {\em Journal of Machine Learning Research}, 3:1137--1155, 2003.

\bibitem{sgd}
L.~Bottou.
\newblock Stochastic gradient learning in neural networks.
\newblock In {\em Proceedings of Neuro-N\^imes 91}, Nimes, France, 1991. EC2.

\bibitem{chandola2009anomaly}
V.~Chandola, A.~Banerjee, and V.~Kumar.
\newblock Anomaly detection: A survey.
\newblock {\em ACM Computing Surveys (CSUR)}, 41(3):15, 2009.

\bibitem{senna1}
R.~Collobert and J.~Weston.
\newblock A unified architecture for natural language processing: Deep neural
  networks with multitask learning.
\newblock In {\em Proceedings of the 25th international conference on Machine
  learning}, pages 160--167. ACM, 2008.

\bibitem{speech1}
G.~E. Dahl, D.~Yu, L.~Deng, and A.~Acero.
\newblock Context-dependent pre-trained deep neural networks for
  large-vocabulary speech recognition.
\newblock {\em Audio, Speech, and Language Processing, IEEE Transactions on},
  20(1):30--42, 2012.

\bibitem{largedeep}
J.~Dean, G.~Corrado, R.~Monga, K.~Chen, M.~Devin, Q.~Le, M.~Mao, M.~Ranzato,
  A.~Senior, P.~Tucker, K.~Yang, and A.~Ng.
\newblock Large scale distributed deep networks.
\newblock In P.~Bartlett, F.~Pereira, C.~Burges, L.~Bottou, and K.~Weinberger,
  editors, {\em Advances in Neural Information Processing Systems 25}, pages
  1232--1240. 2012.

\bibitem{erhanhelps}
D.~Erhan, Y.~Bengio, A.~Courville, P.-A. Manzagol, P.~Vincent, and S.~Bengio.
\newblock Why does unsupervised pre-training help deep learning?
\newblock {\em The Journal of Machine Learning Research}, 11:625--660, 2010.

\bibitem{REF08a}
R.-E. Fan, K.-W. Chang, C.-J. Hsieh, X.-R. Wang, and C.-J. Lin.
\newblock {LIBLINEAR}: A library for large linear classification.
\newblock {\em Journal of Machine Learning Research}, 9:1871--1874, 2008.

\bibitem{fouss2007random}
F.~Fouss, A.~Pirotte, J.-M. Renders, and M.~Saerens.
\newblock Random-walk computation of similarities between nodes of a graph with
  application to collaborative recommendation.
\newblock {\em Knowledge and Data Engineering, IEEE Transactions on},
  19(3):355--369, 2007.

\bibitem{gallagher2010leveraging}
B.~Gallagher and T.~Eliassi-Rad.
\newblock Leveraging label-independent features for classification in sparsely
  labeled networks: An empirical study.
\newblock In {\em Advances in Social Network Mining and Analysis}, pages 1--19.
  Springer, 2010.

\bibitem{Gallagher:2008:UGE:1401890.1401925}
B.~Gallagher, H.~Tong, T.~Eliassi-Rad, and C.~Faloutsos.
\newblock Using ghost edges for classification in sparsely labeled networks.
\newblock In {\em Proceedings of the 14th ACM SIGKDD International Conference
  on Knowledge Discovery and Data Mining}, KDD '08, pages 256--264, New York,
  NY, USA, 2008. ACM.

\bibitem{geman1984stochastic}
S.~Geman and D.~Geman.
\newblock Stochastic relaxation, gibbs distributions, and the bayesian
  restoration of images.
\newblock {\em Pattern Analysis and Machine Intelligence, IEEE Transactions
  on}, (6):721--741, 1984.

\bibitem{getoor2007introduction}
L.~Getoor and B.~Taskar.
\newblock {\em Introduction to statistical relational learning}.
\newblock MIT press, 2007.

\bibitem{Henderson:2011:YKG:2020408.2020512}
K.~Henderson, B.~Gallagher, L.~Li, L.~Akoglu, T.~Eliassi-Rad, H.~Tong, and
  C.~Faloutsos.
\newblock It's who you know: Graph mining using recursive structural features.
\newblock In {\em Proceedings of the 17th ACM SIGKDD International Conference
  on Knowledge Discovery and Data Mining}, KDD '11, pages 663--671, New York,
  NY, USA, 2011. ACM.

\bibitem{distributed}
G.~E. Hinton.
\newblock Learning distributed representations of concepts.
\newblock In {\em Proceedings of the eighth annual conference of the cognitive
  science society}, pages 1--12. Amherst, MA, 1986.

\bibitem{hummel1983foundations}
R.~A. Hummel and S.~W. Zucker.
\newblock On the foundations of relaxation labeling processes.
\newblock {\em Pattern Analysis and Machine Intelligence, IEEE Transactions
  on}, (3):267--287, 1983.

\bibitem{kang2012fast}
U.~Kang, H.~Tong, and J.~Sun.
\newblock Fast random walk graph kernel.
\newblock In {\em SDM}, pages 828--838, 2012.

\bibitem{kondor2002diffusion}
R.~I. Kondor and J.~Lafferty.
\newblock Diffusion kernels on graphs and other discrete input spaces.
\newblock In {\em ICML}, volume~2, pages 315--322, 2002.

\bibitem{vision1}
A.~Krizhevsky, I.~Sutskever, and G.~E. Hinton.
\newblock Imagenet classification with deep convolutional neural networks.
\newblock In {\em NIPS}, volume~1, page~4, 2012.

\bibitem{liben2007link}
D.~Liben-Nowell and J.~Kleinberg.
\newblock The link-prediction problem for social networks.
\newblock {\em Journal of the American society for information science and
  technology}, 58(7):1019--1031, 2007.

\bibitem{cohenASOM}
F.~Lin and W.~Cohen.
\newblock Semi-supervised classification of network data using very few labels.
\newblock In {\em Advances in Social Networks Analysis and Mining (ASONAM),
  2010 International Conference on}, pages 192--199, Aug 2010.

\bibitem{Macskassy03asimple}
S.~A. Macskassy and F.~Provost.
\newblock A simple relational classifier.
\newblock In {\em Proceedings of the Second Workshop on Multi-Relational Data
  Mining (MRDM-2003) at KDD-2003}, pages 64--76, 2003.

\bibitem{macskassy2007classification}
S.~A. Macskassy and F.~Provost.
\newblock Classification in networked data: A toolkit and a univariate case
  study.
\newblock {\em The Journal of Machine Learning Research}, 8:935--983, 2007.

\bibitem{word2vec1}
T.~Mikolov, K.~Chen, G.~Corrado, and J.~Dean.
\newblock Efficient estimation of word representations in vector space.
\newblock {\em CoRR}, abs/1301.3781, 2013.

\bibitem{word2vec2}
T.~Mikolov, I.~Sutskever, K.~Chen, G.~S. Corrado, and J.~Dean.
\newblock Distributed representations of words and phrases and their
  compositionality.
\newblock In {\em Advances in Neural Information Processing Systems 26}, pages
  3111--3119. 2013.

\bibitem{regularities}
T.~Mikolov, W.-t. Yih, and G.~Zweig.
\newblock Linguistic regularities in continuous space word representations.
\newblock In {\em Proceedings of NAACL-HLT}, pages 746--751, 2013.

\bibitem{hsm2}
A.~Mnih and G.~E. Hinton.
\newblock A scalable hierarchical distributed language model.
\newblock {\em Advances in neural information processing systems},
  21:1081--1088, 2009.

\bibitem{hsm1}
F.~Morin and Y.~Bengio.
\newblock Hierarchical probabilistic neural network language model.
\newblock In {\em Proceedings of the international workshop on artificial
  intelligence and statistics}, pages 246--252, 2005.

\bibitem{neville2000iterative}
J.~Neville and D.~Jensen.
\newblock Iterative classification in relational data.
\newblock In {\em Proc. AAAI-2000 Workshop on Learning Statistical Models from
  Relational Data}, pages 13--20, 2000.

\bibitem{Neville:2005:LRA:1090193.1090201}
J.~Neville and D.~Jensen.
\newblock Leveraging relational autocorrelation with latent group models.
\newblock In {\em Proceedings of the 4th International Workshop on
  Multi-relational Mining}, MRDM '05, pages 49--55, New York, NY, USA, 2005.
  ACM.

\bibitem{neville2008bias}
J.~Neville and D.~Jensen.
\newblock A bias/variance decomposition for models using collective inference.
\newblock {\em Machine Learning}, 73(1):87--106, 2008.

\bibitem{newman2006modularity}
M.~E. Newman.
\newblock Modularity and community structure in networks.
\newblock {\em Proceedings of the National Academy of Sciences},
  103(23):8577--8582, 2006.

\bibitem{Pearl:1988:PRI:534975}
J.~Pearl.
\newblock {\em Probabilistic Reasoning in Intelligent Systems: Networks of
  Plausible Inference}.
\newblock Morgan Kaufmann Publishers Inc., San Francisco, CA, USA, 1988.

\bibitem{hogwild}
B.~Recht, C.~Re, S.~Wright, and F.~Niu.
\newblock Hogwild: A lock-free approach to parallelizing stochastic gradient
  descent.
\newblock In {\em Advances in Neural Information Processing Systems 24}, pages
  693--701. 2011.

\bibitem{sen2008collective}
P.~Sen, G.~Namata, M.~Bilgic, L.~Getoor, B.~Galligher, and T.~Eliassi-Rad.
\newblock Collective classification in network data.
\newblock {\em AI magazine}, 29(3):93, 2008.

\bibitem{spielman2004nearly}
D.~A. Spielman and S.-H. Teng.
\newblock Nearly-linear time algorithms for graph partitioning, graph
  sparsification, and solving linear systems.
\newblock In {\em Proceedings of the thirty-sixth annual ACM symposium on
  Theory of computing}, pages 81--90. ACM, 2004.

\bibitem{Tang:2009:RLV:1557019.1557109}
L.~Tang and H.~Liu.
\newblock Relational learning via latent social dimensions.
\newblock In {\em Proceedings of the 15th ACM SIGKDD International Conference
  on Knowledge Discovery and Data Mining}, KDD '09, pages 817--826, New York,
  NY, USA, 2009. ACM.

\bibitem{tang2009scalable}
L.~Tang and H.~Liu.
\newblock Scalable learning of collective behavior based on sparse social
  dimensions.
\newblock In {\em Proceedings of the 18th ACM conference on Information and
  knowledge management}, pages 1107--1116. ACM, 2009.

\bibitem{Tang:2011:Leveraging}
L.~Tang and H.~Liu.
\newblock Leveraging social media networks for classification.
\newblock {\em Data Mining and Knowledge Discovery}, 23(3):447--478, 2011.

\bibitem{vishwanathan2010graph}
S.~Vishwanathan, N.~N. Schraudolph, R.~Kondor, and K.~M. Borgwardt.
\newblock Graph kernels.
\newblock {\em The Journal of Machine Learning Research}, 99:1201--1242, 2010.

\bibitem{wang2013multi}
X.~Wang and G.~Sukthankar.
\newblock Multi-label relational neighbor classification using social context
  features.
\newblock In {\em Proceedings of the 19th ACM SIGKDD international conference
  on Knowledge discovery and data mining}, pages 464--472. ACM, 2013.

\bibitem{zachary1977information}
W.~Zachary.
\newblock An information flow modelfor conflict and fission in small groups1.
\newblock {\em Journal of anthropological research}, 33(4):452--473, 1977.

\end{thebibliography}
